\documentclass[twocolumn,showpacs,preprintnumbers,superscriptaddress,amsmath,amssymb,nofootinbib]{revtex4}

\usepackage{graphicx}
\usepackage{bm}

\usepackage[normalem]{ulem}  
\usepackage[dvipdfmx]{color} 

\renewcommand\sout{\bgroup \color{red} \ULdepth=-.5ex \ULset}

\newcommand{\Psfig}[2]{\includegraphics[width=#1]{#2}}
\newcommand{\PsfigII}[2]{\includegraphics[scale=#1]{#2}}

\def\fzero{f_{0}(980)}
\def\azero{a_{0}(980)}
\def\afmix{$\azero$-$\fzero$ }

\def\mev{\text{ MeV}}
\def\gev{\text{ GeV}}
\def\fm{\text{ fm}}

\def\phph{\phantom{-}}

\allowdisplaybreaks[1]

\begin{document}

\preprint{}

\title{Constraint on $\bm{K \bar{K}}$ compositeness of the
  $\bm{a_{0}(980)}$ and $\bm{f_{0}(980)}$ resonances from their mixing
  intensity}

\author{T.~Sekihara} 
\email{sekihara@rcnp.osaka-u.ac.jp}
\affiliation{Research Center for Nuclear Physics
  (RCNP), Osaka University, Ibaraki, Osaka, 567-0047, Japan}

\author{S.~Kumano}
\affiliation{KEK Theory Center, Institute of Particle and Nuclear
  Studies, High Energy Accelerator Research Organization (KEK), 1-1,
  Oho, Tsukuba, Ibaraki 305-0801, Japan}
\affiliation{J-PARC Branch, KEK Theory Center,
  Institute of Particle and Nuclear Studies, 
  High Energy Accelerator Research Organization (KEK),
  203-1, Shirakata, Tokai, Ibaraki, 319-1106, Japan}

\date{\today}

\begin{abstract}
  Structure of the $a_{0}(980)$ and $f_{0}(980)$ resonances is
  investigated with the $a_{0}(980)$-$f_{0}(980)$ mixing intensity
  from the viewpoint of compositeness, which corresponds to the amount
  of two-body states composing resonances as well as bound states.
  For this purpose we first formulate the $a_{0}(980)$-$f_{0}(980)$
  mixing intensity as the ratio of two partial decay widths of a
  parent particle, in the same manner as the recent analysis in BES
  experiments.  Calculating the $a_{0}(980)$-$f_{0}(980)$ mixing
  intensity with the existing Flatte parameters from experiments, we
  find that many combinations of the $a_{0}(980)$ and $f_{0}(980)$
  Flatte parameters can reproduce the experimental value of the
  $a_{0}(980)$-$f_{0}(980)$ mixing intensity by BES.  Next, from the
  same Flatte parameters we also calculate the $K \bar{K}$
  compositeness for $a_{0}(980)$ and $f_{0}(980)$.  Although the
  compositeness with the correct normalization becomes complex in
  general for resonance states, we find that the Flatte parameters for
  $f_{0}(980)$ imply large absolute value of the $K \bar{K}$
  compositeness and the parameters for $a_{0}(980)$ lead to small but
  nonnegligible absolute value of the $K \bar{K}$ compositeness.
  Then, connecting the mixing intensity and the $K \bar{K}$
  compositeness via the $a_{0}(980)$- and $f_{0}(980)$-$K \bar{K}$
  coupling constants, we establish a relation between them.  As a
  result, a small mixing intensity indicates a small value of the
  product of the $K \bar{K}$ compositeness for the $a_{0}(980)$ and
  $f_{0}(980)$ resonances.  Moreover, the experimental value of the
  $a_{0}(980)$-$f_{0}(980)$ mixing intensity implies that the
  $a_{0}(980)$ and $f_{0}(980)$ resonances cannot be simultaneously $K
  \bar{K}$ molecular states.
\end{abstract}

\pacs{%
  14.40.Be,  
  12.39.Mk,  
  21.45.-v  
}
\maketitle

\section{Introduction}

The nature of the lightest scalar meson nonet [$f_{0}(500)$ or
$\sigma$, $K_{0}^{\ast}(800)$ or $\kappa$, $f_{0}(980)$, and
$a_{0}(980)$] has been a hot topic in hadron physics for many
years~\cite{Olive:1900zz}.  A na\"{i}ve expectation with the
$q\bar{q}$ configuration indicates that they should show the same mass
ordering as, {\it e.g.}, the vector meson nonet, but in real they
exhibit inverted spectrum from the expectation.  For this reason they
have been considered to be exotic hadrons, which are not able to be
classified as $q\bar{q}$ for mesons and $qqq$ for baryons.  Indeed, in
Refs.~\cite{Jaffe:1976ig, Jaffe:1976ih} it was suggested that in a bag
model the interaction between quarks inside a compact $q q \bar{q}
\bar{q}$ system is attractive especially in the scalar channel and
hence the light scalar mesons would be compact $q q \bar{q} \bar{q}$
systems.  However, it was found that in a nonrelativistic quark model
$K \bar{K}$ molecules can appear as weakly bound $s$-wave states,
which may be identified with $f_{0}(980)$ and
$a_{0}(980)$~\cite{Weinstein:1982gc, Weinstein:1983gd}.  The lightest
scalar mesons can also be described by the combination of the chiral
perturbation theory and the scattering unitarity~\cite{Truong:1988zp,
  Truong:1991gv, Dobado:1989qm, Dobado:1993ha, Dobado:1996ps,
  Oller:1997ti, Oller:1997ng, Oller:1998hw, Oller:1998zr,
  GomezNicola:2001as} in pseudoscalar meson-pseudoscalar meson
scatterings from the hadronic degrees of freedom.  This fact implies
that the lightest scalar mesons may have nonnegligible components of
hadronic molecules.  In a model-independent way, on the other hand,
the structure of $f_{0}(980)$ and $a_{0}(980)$ was discussed in
Ref.~\cite{Baru:2003qq}, which suggested that $f_{0}(980)$ should be a
$K \bar{K}$ molecular state to a large degree and $a_{0}(980)$ also 
seems to have a nonnegligible $K \bar{K}$ component.  There are
further discussions on their structure as well, {\it e.g.}, hybrid
states for $f_{0}(980)$ and $a_{0}(980)$~\cite{Ishida:1995km}.

\begin{figure*}[!ht]
  \centering
  \begin{tabular*}{\textwidth}{@{\extracolsep{\fill}}ccc}
    \PsfigII{0.24}{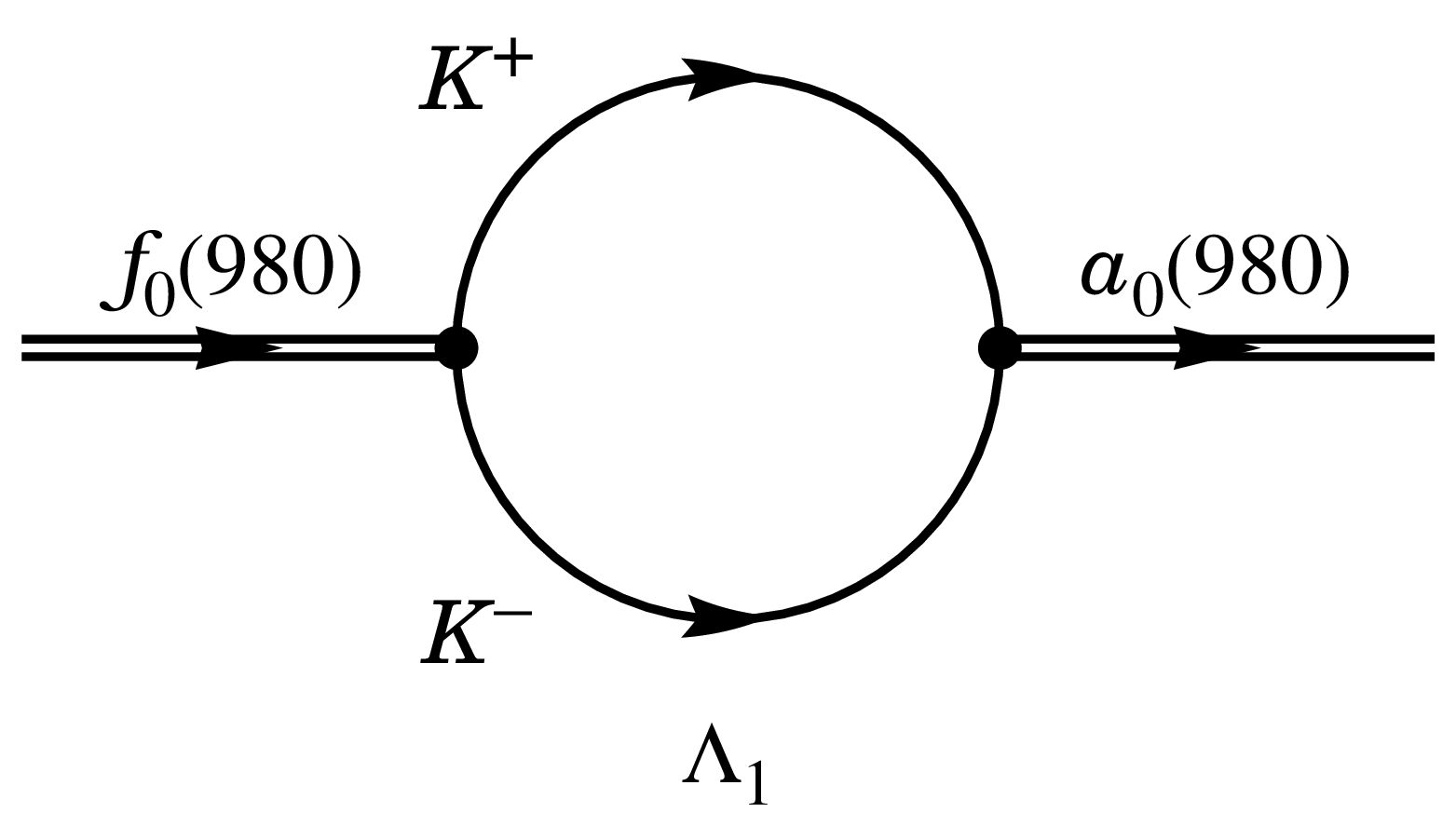} & 
    \PsfigII{0.24}{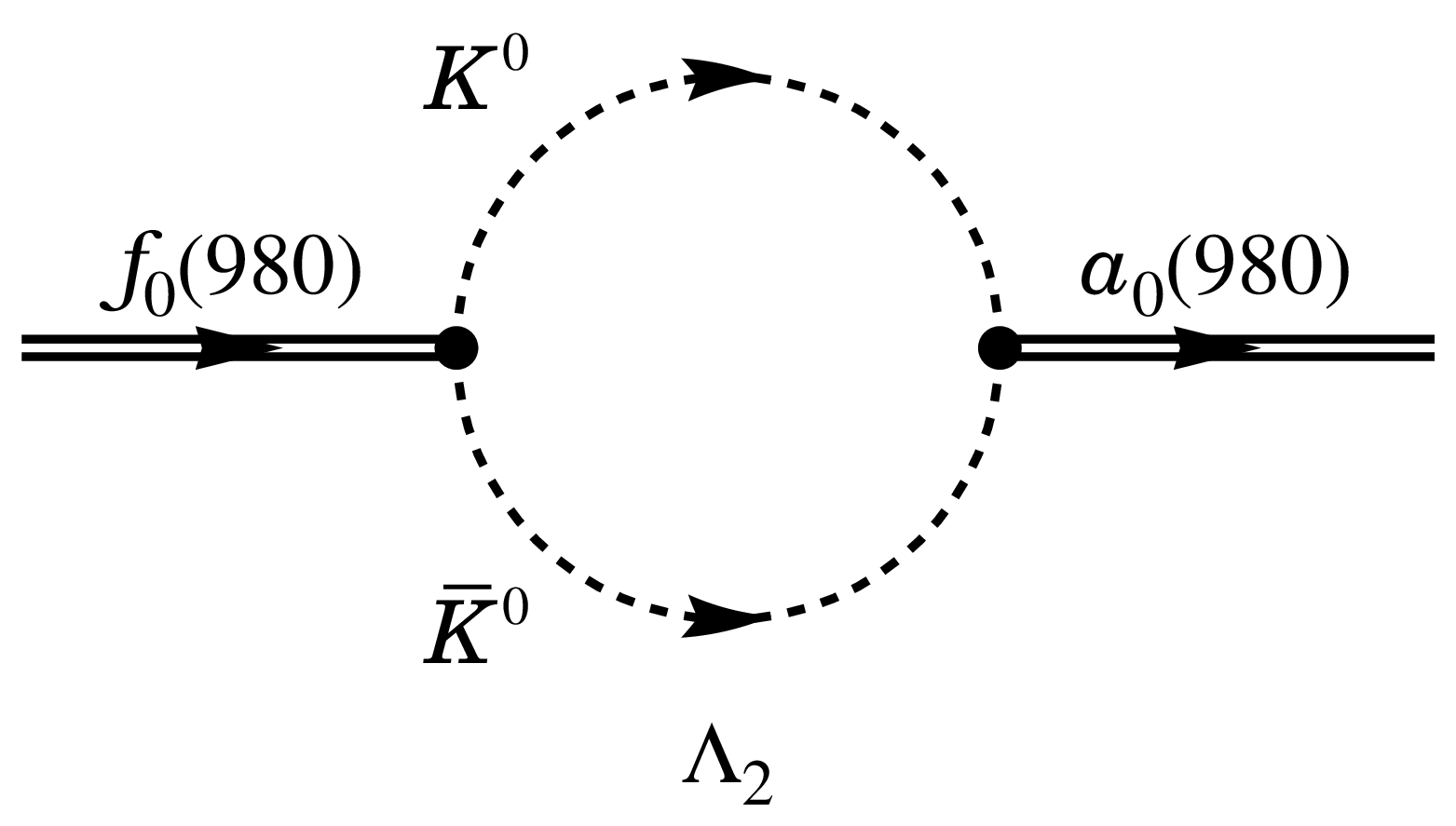} & 
    \PsfigII{0.24}{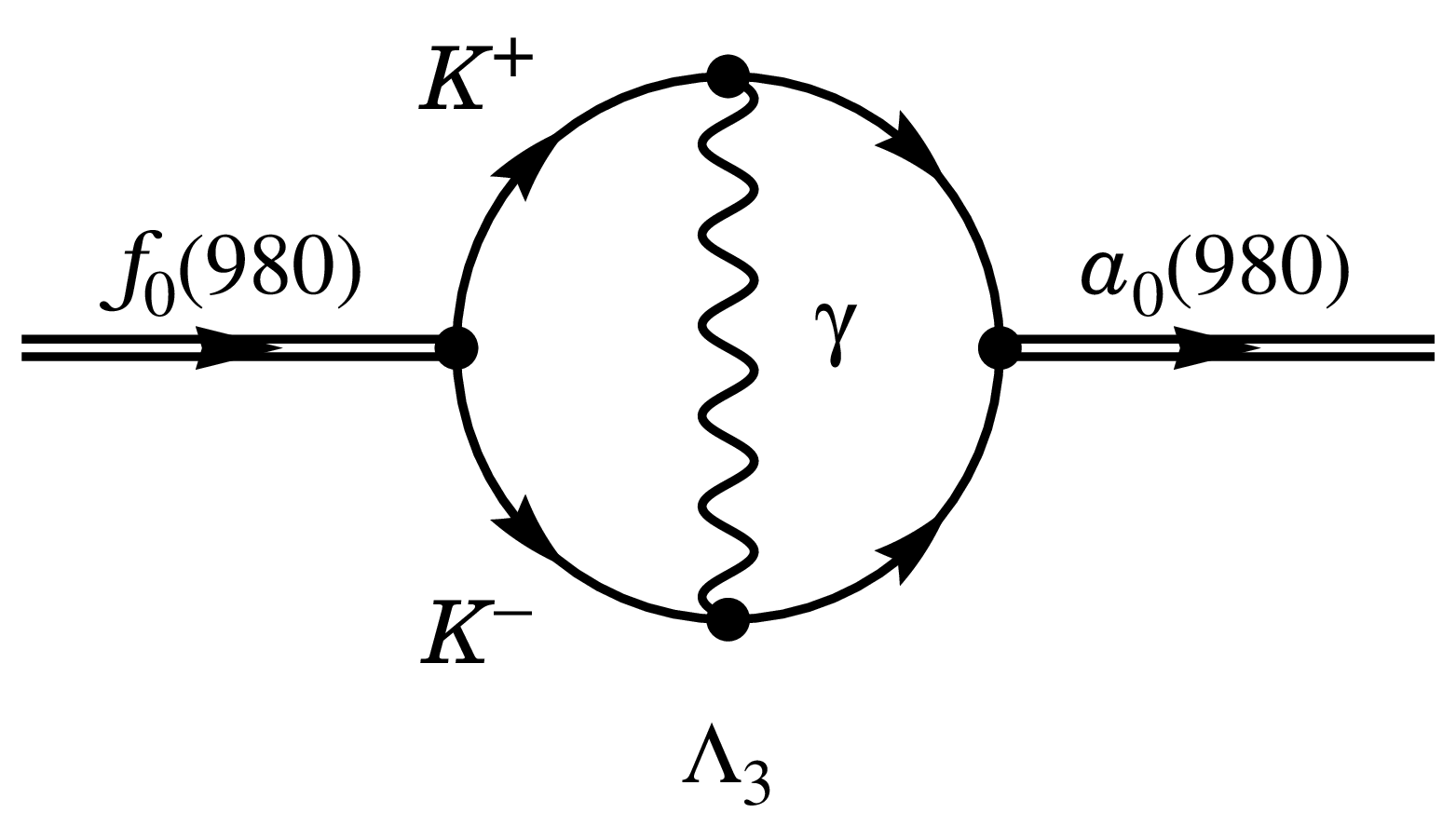} 
  \end{tabular*}
  \caption{Feynman diagrams for the leading contribution ($\Lambda
    _{1} + \Lambda _{2}$) and a subleading contribution ($\Lambda
    _{3}$) to the $a_{0}(980)$-$f_{0}(980)$ mixing.}
  \label{fig:1}
\end{figure*}

Among the light scalar mesons, $a_{0}(980)$ and $f_{0}(980)$ are of
special interest because their almost degenerate masses would lead to
a mixing of these mesons in isospin symmetry violating processes.  In
particular, it was pointed out in Ref.~\cite{Achasov:1979xc} that the
difference of the unitarity cuts between the charged and neutral $K
\bar{K}$ pairs, which thresholds are close to the $a_{0}(980)$ and
$f_{0}(980)$ masses, can enhance the $a_{0}(980)$-$f_{0}(980)$ mixing
to be sizable compared to, {\it e.g.}, the $\rho (770)$-$\omega (782)$
mixing.  Namely, the leading contribution to the
$a_{0}(980)$-$f_{0}(980)$ mixing comes from the mixing amplitude
$\Lambda _{1} + \Lambda _{2}$ in Fig.~\ref{fig:1}, and it behaves
\begin{equation}
\Lambda _{1} + \Lambda _{2} 
= \mathcal{O} ( p_{K^{0}} - p_{K^{+}} ) ,
\end{equation}
where $p_{K^{0}}$ ($p_{K^{+}}$) denotes the magnitude of the relative
momentum of the neutral (charged) kaon pair.  Then, due to the
difference of the unitary cuts, the mixing effect should be unusually
enhanced in the energy between $m_{K^{+}} + m_{K^{-}} = 987 \mev$ and
$m_{K^{0}} + m_{\bar{K}^{0}} = 995 \mev$ to be
\begin{equation}
\Lambda _{1} + \Lambda _{2} 
= \mathcal{O} \left ( \sqrt{\frac{m_{K^{0}}^{2} - m_{K^{+}}^{2}}
{m_{K^{0}}^{2} + m_{K^{+}}^{2}}} \right ) ,
\end{equation}
while out of the energy region the mixing effect returns to a value of
natural size, $\mathcal{O} [ ( m_{K^{0}}^{2} - m_{K^{+}}^{2} ) / (
  m_{K^{0}}^{2} + m_{K^{+}}^{2} )]$.  In addition, as a subleading
contribution the electromagnetic interaction would enhance the \afmix
mixing, since the electromagnetic interaction takes place selectively
in the $K^{+} K^{-}$ loop.  Bearing in mind that in general a scalar
meson does not have derivative couplings to two pseudoscalar mesons,
we have only a soft photon exchange between $K^{+}$ and $K^{-}$ as the
leading order with respect to the electromagnetic interaction, which
is diagrammatically shown as $\Lambda _{3}$ in Fig.~\ref{fig:1}.
Indeed, the amplitude $\Lambda _{3}$ logarithmically diverges at the
$K^{+} K^{-}$ threshold in an approximation of the threshold
expansion.  For observations of the $a_{0}(980)$-$f_{0}(980)$ mixing,
various reactions which should be sensitive to the mixing were
discussed in, {\it e.g.}, Refs.~\cite{Kerbikov:2000pu, Close:2000ah,
  Achasov:2002hg, Achasov:2003se, Wu:2007jh, Hanhart:2007bd,
  Wu:2008hx, Aceti:2012dj, Roca:2012cv}, and the mixing effect was
recently observed in an experiment~\cite{Ablikim:2010aa} from the
decay of $J/\psi$.

The $a_{0}(980)$-$f_{0}(980)$ mixing has been expected to shed light
on the structure of the $a_{0}(980)$ and $f_{0}(980)$ resonances.
Actually in Ref.~\cite{Ablikim:2010aa} the experimental value of the
mixing intensity was compared to several theoretical predictions and
structure of the two resonances was discussed.  We here emphasize that
coupling constants of $a_{0}(980)$-$K \bar{K}$ and $f_{0}(980)$-$K
\bar{K}$ reflect the $K \bar{K}$ structure of the $a_{0}(980)$ and
$f_{0}(980)$ resonances, respectively; especially a larger $K \bar{K}$
coupling constant means a larger fraction of the $K \bar{K}$ component
in the scalar mesons~\cite{Hanhart:2007cm}.  In recent studies this
statement has been formulated in terms of
compositeness~\cite{Hyodo:2011qc, Aceti:2012dd, Xiao:2012vv,
  Aceti:2013jg, Hyodo:2013nka, Aceti:2014wka, Sekihara:2014} in the
so-called chiral unitary approach, which is a way to combine the
chiral perturbation theory and the scattering unitarity.  In these
studies the compositeness was defined as the two-body composite part
of the normalization of the total wave function, and hence the
compositeness corresponds to the amount of the two-body states
composing a resonance as well as a bound state.  In the formulation
the two-body wave function was found to be proportional to the
coupling constant of the resonance state to the two-body
state~\cite{Sekihara:2014, Gamermann:2009uq, YamagataSekihara:2010pj}.
Thus, bearing in mind that the $a_{0}(980)$-$f_{0}(980)$ mixing
amplitude contains both the coupling constants of $a_{0}(980)$-$K
\bar{K}$ and $f_{0}(980)$-$K \bar{K}$, one can expect a relation
between the $K \bar{K}$ compositeness of the $a_{0}(980)$ and
$f_{0}(980)$ resonances and their mixing intensity through the
strength of the coupling constants of $a_{0}(980)$-$K \bar{K}$ and
$f_{0}(980)$-$K \bar{K}$ in the mixing amplitude, in a similar manner
to the relation between the $\Lambda (1405)$ radiative decay width and
its $\bar{K} N$ compositeness established in
Ref.~\cite{Sekihara:2013sma}.  The purpose of this paper is to
establish a relation between the $K \bar{K}$ compositeness of the
$a_{0}(980)$ and $f_{0}(980)$ resonances and their mixing intensity
and to give a constraint on the structure of the two resonances from
the experimental value of the mixing intensity obtained in
Ref.~\cite{Ablikim:2010aa}.

This paper is organized as follows.  In Sec.~\ref{sec:form} we
formulate the $a_{0}(980)$-$f_{0}(980)$ mixing intensity.  In this
section we also calculate the $a_{0}(980)$-$f_{0}(980)$ mixing
intensity with several Flatte parameter sets for $a_{0}(980)$ and
$f_{0}(980)$ from experiments, and compare the numerical results with
the recent experimental result.  Next, in Sec.~\ref{sec:comp} we
develop our formulation of the compositeness in the context of the
chiral unitary approach, and we calculate the $K \bar{K}$
compositeness of $a_{0}(980)$ and $f_{0}(980)$ with the experimental
Flatte parameter sets.  Then, in Sec.~\ref{sec:relation} we give a
relation between the mixing intensity and the $K \bar{K}$
compositeness for the $a_{0}(980)$ and $f_{0}(980)$ resonances.
Moreover, we discuss further steps for the determination of the
structure of the $a_{0}(980)$ and $f_{0}(980)$ resonances in
Sec.~\ref{sec:discussion}.  Section~\ref{sec:conclusion} is devoted to
the conclusion of this study.

\section{The $\bm{a_{0}(980)}$-$\bm{f_{0}(980)}$ mixing intensity}
\label{sec:form}

In this section we formulate the $a_{0}(980)$-$f_{0}(980)$ mixing
intensity.  For this purpose we first determine the expression of the
$a_{0}(980) \leftrightarrow f_{0}(980)$ mixing amplitude in
Sec.~\ref{sec:form-1}.  Next we evaluate the propagators of
$a_{0}(980)$ and $f_{0}(980)$ with their mixing in
Sec.~\ref{sec:form-2}.  Then we formulate the
$a_{0}(980)$-$f_{0}(980)$ mixing intensity as the ratio of partial
decay widths of a parent particle in Sec.~\ref{sec:form-3}.  Finally
in Sec.~\ref{sec:mixing} we calculate the mixing intensity by using
several parameter sets obtained from experimental data.

\subsection{Mixing amplitude}
\label{sec:form-1}

First of all we determine the $a_{0}(980) \leftrightarrow f_{0}(980)$
mixing amplitude $\Lambda (s)$ as a function of the squared
momentum of the scalar mesons, $s$.  In this study we consider three
Feynman diagrams in Fig.~\ref{fig:1} for the $a_{0}(980)
\leftrightarrow f_{0}(980)$ mixing, and the mixing amplitude
$\Lambda (s)$ is sum of the three contributions:
\begin{equation}
\Lambda ( s ) = 
\Lambda _{1} ( s ) + \Lambda _{2} ( s ) + \Lambda _{3} ( s ) .
\label{eq:mix_amp}
\end{equation}
Here we assume isospin symmetry for coupling constants.  Namely,
$a_{0}(980)$-$K \bar{K}$ and $f_{0}(980)$-$K \bar{K}$ coupling
constants in the particle basis, $\bar{g}_{a}$ and $\bar{g}_{f}$, are
given as\footnote{We put bar on the coupling constants, $\bar{g}_{a,
    f}$, which are used on the real energy axis.  On the other hand,
  we will not put bar on the coupling constants which are evaluated as
  the residue of the scattering amplitude (see Sec.~\ref{sec:comp}). }
\begin{equation}
\bar{g}_{a} = \bar{g}_{a K^{+} K^{-}} = - \bar{g}_{a K^{0} \bar{K}^{0}} , 
\quad 
\bar{g}_{f} = \bar{g}_{f K^{+} K^{-}} = \bar{g}_{f K^{0} \bar{K}^{0}} . 
\label{eq:gKK_PB}
\end{equation}
Then it was pointed out in Ref.~\cite{Achasov:1979xc} that the sum of
the first and second contributions, $\Lambda _{1} + \Lambda _{2}$,
converges and the result can be presented as an expansion in the $K
\bar{K}$ phase space:
\begin{align}
\Lambda _{1} ( s ) + \Lambda _{2} ( s ) 
= & - \frac{i}{16 \pi} 
\bar{g}_{a} \bar{g}_{f} 
[ \sigma _{1} ( s ) - \sigma _{2} ( s ) ]
\nonumber \\ & 
+ \mathcal{O} [ \sigma _{1}^{2} ( s ) - \sigma _{2}^{2} ( s ) ], 
\label{eq:Lambda-PS-sym}
\end{align}
where $i=1$ ($2$) denotes the channel $K^{+} K^{-}$ ($K^{0}
\bar{K}^{0}$) and the phase space $\sigma _{i}(s)$ is defined as
\begin{equation}
\sigma _{i} ( s ) \equiv 
\frac{\lambda ^{1/2} ( s, \, m_{i}^{2}, \, m_{i}^{2})}{s} = 
\sqrt{1 - \frac{4 m_{i}^{2}}{s}} , 
\quad 
i = 1, \, 2 ,
\end{equation}
with the K\"{a}llen function $\lambda (x, \, y, \, z) = x^{2} + y^{2}
+ z^{2} - 2 x y - 2 y z - 2 z x$ and masses $m_{1} = m_{K^{+}}$ and
$m_{2} = m_{K^{0}}$.\footnote{In our calculations we use the physical
  masses $m_{K^{+}} = m_{K^{-}} = 493.68 \mev$, $m_{K^{0}} =
  m_{\bar{K}^{0}} = 497.61 \mev$, and $m_{\eta} = 547.85 \mev$, while
  for the isospin symmetric masses we use $m_{\pi} = (m_{\pi ^{+}} +
  m_{\pi ^{-}} + m_{\pi ^{0}} ) / 3 = 138.04 \mev$ and $m_{K} =
  (m_{K^{+}} + m_{K^{-}} + m_{K^{0}} + m_{\bar{K}^{0}} ) / 4 = 495.65
  \mev$.}  Since we have taken into account just the difference of the
unitary cut contributions, this leading-order contribution is model
independent except for the coupling constants.

The third contribution to the mixing amplitude, $\Lambda _{3}$, is a
soft photon-exchange diagram between $K^{+} K^{-}$, and with the 
photon-exchange loop function $G_{\gamma} (s)$ the mixing amplitude 
can be written as 
\begin{equation}
\Lambda _{3} ( s ) = \bar{g}_{a}
G_{\gamma} ( s ) \bar{g}_{f} . 
\end{equation}
For the evaluation of the photon-exchange loop function $G_{\gamma}
(s)$, we take an approximation by the threshold-expanded
form~\cite{Beneke:1997zp}, which reads~\cite{Hanhart:2007bd}
\begin{align}
G_{\gamma} ( s ) = 
& - \frac{\alpha}{32 \pi}
\left [ \ln \frac{4 m_{K^{+}}^{2} - s}{m_{K^{+}}^{2}} 
+ \ln 2 + \frac{21 \zeta (3)}{2 \pi ^{2}} \right ]
\nonumber \\
& + \mathcal{O} [ ( s - 4 m_{K^{+}}^{2} )^{2} ] , 
\end{align}
with the fine structure constant $\alpha \approx 1/137$ and the zeta
function $\zeta (x)$ with $\zeta (3) = 1.20205 \dots$.

In above expressions, only the two coupling constants, $\bar{g}_{a}$
and $\bar{g}_{f}$, are the parameters and reflect the structure of the
$a_{0}(980)$ and $f_{0}(980)$ resonances.  In this study the coupling
constants are taken from the Flatte parameter sets with several
experimental fittings in Sec.~\ref{sec:mixing}, and then in
Sec.~\ref{sec:relation} they are used to establish a relation between
the mixing intensity and the $K \bar{K}$ compositeness of the
$a_{0}(980)$ and $f_{0}(980)$ resonances.

\subsection{Propagators of $\bm{a_{0}(980)}$ and $\bm{f_{0}(980)}$ 
with their mixing}
\label{sec:form-2}

Next we formulate propagators of the $a_{0}(980)$ and $f_{0}(980)$
mesons with their mixing.  If the $a_{0}(980)$-$f_{0}(980)$ mixing is
absent, their propagators can be expressed as $1/D_{a}(s)$ and
$1/D_{f}(s)$ in the Flatte parametrization~\cite{Flatte:1976xu}
\begin{equation}
  \begin{split}
    & D_{a} ( s ) \equiv s - M_{a}^{2} + i \sqrt{s} 
    [ \Gamma _{\pi \eta}^{a} ( s ) + \Gamma _{K \bar{K}}^{a} ( s ) ] , 
    \\
    & D_{f} ( s ) \equiv s - M_{f}^{2} + i \sqrt{s} 
    [ \Gamma _{\pi \pi}^{f} ( s ) + \Gamma _{K \bar{K}}^{f} ( s ) ] , 
    \label{eq:Flatte_af}
  \end{split}
\end{equation}
for $a_{0}(980)$ and $f_{0}(980)$, respectively.  Here $s$ is the
squared momentum of the scalar mesons, $M_{a}$ and $M_{f}$ are masses
of $a_{0}(980)$ and $f_{0}(980)$, respectively, and decay width of $a
\to b + c$, $\Gamma _{b c}^{a} (s)$, is defined as
\begin{equation}
\Gamma _{b c}^{a} ( s ) 
\equiv \frac{| \bar{g}_{a b c} |^{2}}{8 \pi s} p_{b c} ( s ) , 
\quad
p_{b c} ( s ) \equiv 
\frac{\lambda ^{1/2} (s, \, m_{b}^{2}, \, m_{c}^{2})}
{2 \sqrt{s}} ,
\label{eq:Gamma_abc}
\end{equation}
with the $a$-$b c$ coupling constant in the isospin basis $\bar{g}_{a
  b c}$, the magnitude of the relative momentum $p_{b c}$, and the $b$
and $c$ masses $m_{b}$ and $m_{c}$, respectively.  Here we note that,
due to the energy dependence of the decay-width terms in $D_{a (f)}$,
the pole position of the propagator slightly shifts from that of the
na\"{i}ve expectation $s = [M_{a (f)} - i \Gamma ^{a (f)} (M_{a
    (f)}^{2})/2]^{2}$.  Furthermore, the momentum $p_{b c} (s)$ in the
decay width~\eqref{eq:Gamma_abc} requires us to move to the proper
Riemann sheet when we search for the pole of the propagator.
Throughout this study we search for the $\azero$ [$\fzero$] pole
existing in the unphysical Riemann sheet of the $\pi \eta$ ($\pi \pi$)
channel and in the physical Riemann sheet of the $K \bar{K}$ channel.
We also note that the isospin symmetry breaking negligibly affects the
decay width $\Gamma _{b c}^{a} (s)$ in this study, so we use the
isospin symmetric masses and coupling constants of pions and kaons for
the evaluation of the decay width~\eqref{eq:Gamma_abc}.

Now let us turn on the $a_{0}(980)$-$f_{0}(980)$ mixing.  In this
condition we can obtain the $a_{0}(980)$ propagator with the
$a_{0}(980)$-$f_{0}(980)$ mixing, $P_{a}(s)$, by summing up all the
contributions of $a_{0}(980) \to f_{0}(980) \to \cdots \to
a_{0}(980)$, and hence $P_{a}(s)$ is expressed as
\begin{align}
P_{a} ( s ) = & \frac{1}{D_{a}} + \frac{1}{D_{a}} \Lambda \frac{1}{D_{f}}
\Lambda \frac{1}{D_{a}} + \cdots 
= \frac{1}{D_{a}} \sum _{n=0}^{\infty} 
\left ( \frac{\Lambda ^{2}}{D_{a} D_{f}} \right ) ^{n} 
\nonumber \\
= & \frac{1}{D_{a}} 
\left ( 1 - \frac{\Lambda ^{2}}{D_{a} D_{f}} \right ) ^{-1}
= \frac{D_{f}}{D_{a} D_{f} - \Lambda ^{2}} ,
\end{align}
where $\Lambda (s)$ is the $a_{0}(980) \leftrightarrow f_{0}(980)$
mixing amplitude determined in the previous subsection.  In
similar manners we can obtain the $f_{0}(980)$ propagator $P_{f}(s)$,
the $a_{0}(980) \to f_{0}(980)$ propagator $P_{a \to f}(s)$, and the
$f_{0}(980) \to a_{0}(980)$ propagator $P_{f \to a}(s)$ with the
$a_{0}(980)$-$f_{0}(980)$ mixing, and they are summarized as follows:
\begin{equation}
  \begin{pmatrix}
    P_{a} & P_{a \to f} \\
    P_{f \to a} & P_{f} \\
  \end{pmatrix} 
  = \frac{1}{D_{a} D_{f} - \Lambda ^{2}} 
  \begin{pmatrix}
    D_{f} & \Lambda \\
    \Lambda & D_{a} \\
  \end{pmatrix} 
  .
  \label{eq:prop_mixing}
\end{equation}

\subsection{Partial decay widths and mixing intensity}
\label{sec:form-3}

\begin{figure}[!t]
  \centering
  \begin{tabular}{c}
    \PsfigII{0.24}{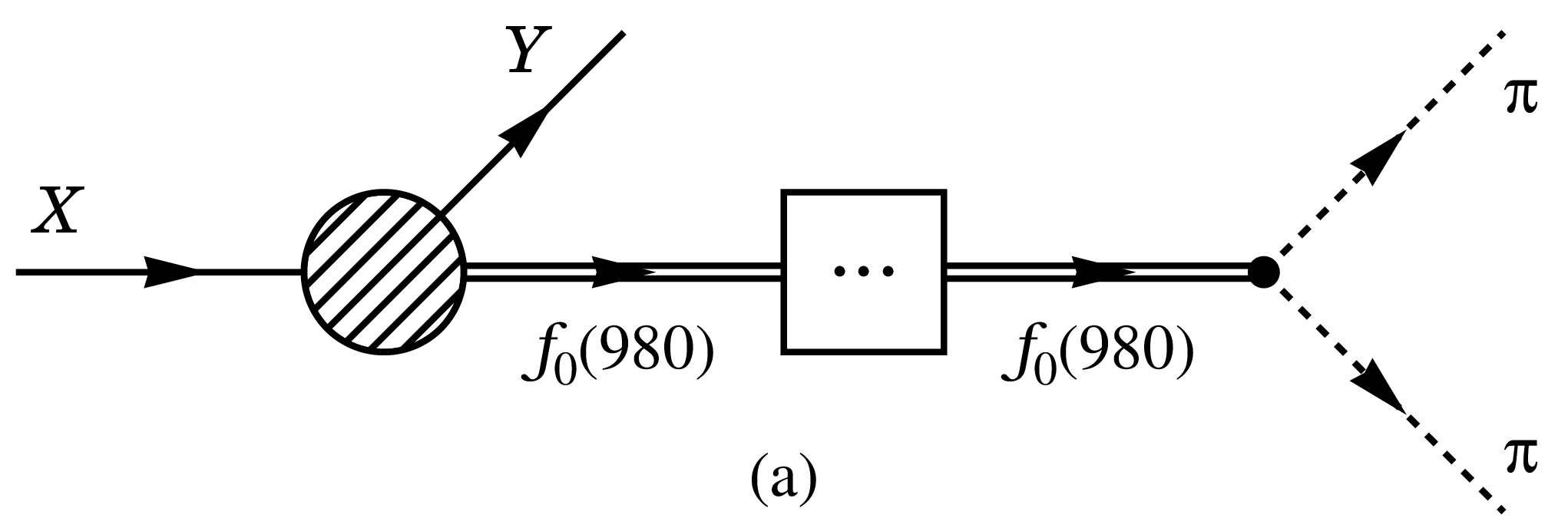} \\
    \PsfigII{0.24}{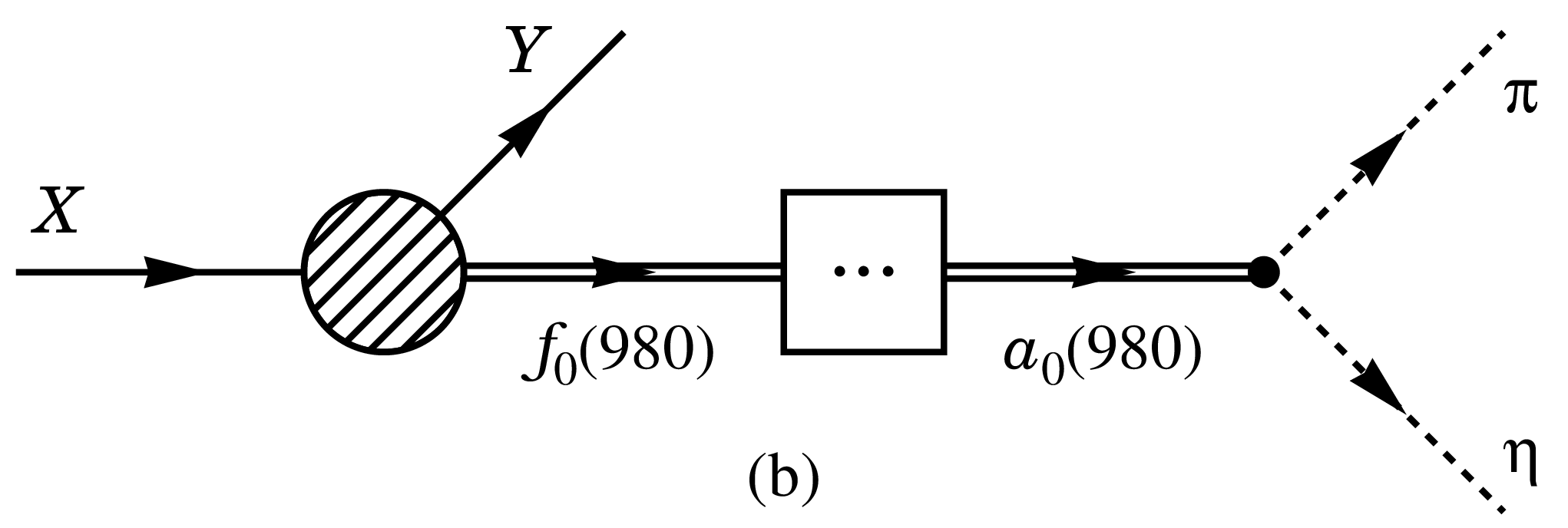} 
  \end{tabular}
  \caption{Schematic diagrams of the decays (a) $X \to Y f_{0}(980)
    \to Y \pi \pi$ and (b) $X \to Y f_{0}(980) \to Y a_{0}(980) \to Y
    \pi \eta$.  In the figures ellipses denote that the propagators
    of the scalar mesons include the $a_{0}(980)$-$f_{0}(980)$ mixing
    contribution.}
 \label{fig:2}
\end{figure}

Let us now define the $a_{0}(980)$-$f_{0}(980)$ mixing intensity $\xi
_{f a}$.  In the experimental analysis in Ref.~\cite{Ablikim:2010aa}
the mixing intensity was defined as the ratio of two branching
fractions of $J/\psi$, $J/\psi \to \phi f_{0}(980) \to \phi a_{0}(980)
\to \phi \pi \eta$ to $J/\psi \to \phi f_{0}(980) \to \phi \pi \pi$.
Hence, in the same manner as this experimental analysis, we define the
$a_{0}(980)$-$f_{0}(980)$ mixing intensity $\xi _{f a}$ as the ratio
of the decay widths:
\begin{equation}
\xi _{f a} \equiv 
\frac{\Gamma _{X, a}}{\Gamma _{X, f}} , 
\label{eq:intensity}
\end{equation}
where $\Gamma _{X, a}$ and $\Gamma _{X, f}$ are partial decay widths
of a meson $X$ to $Y f_{0}(980) \to Y a_{0}(980) \to Y \pi \eta$ and
to $Y f_{0}(980) \to Y \pi \pi$, respectively.  Here we assume that
both $X$ and $Y$ are $I = 0$ states, and hence isospin symmetry allows
only the $X$-$Y$-$\fzero$ vertex.  Therefore, in our formulation
$\azero$ appears only through the \afmix mixing.  We could consider an
intrinsic isospin-violating contribution which allows a direct
coupling of $X$ to the $Y$-$\azero$ system, but such a contribution
will scale as a natural size, for instance, $(m_{d} - m_{u})/m_{s}$,
and will be much smaller than the mixing amplitude of the $K \bar{K}$
loops between the charged and neutral $K \bar{K}$ thresholds.  For
this reason we neglect the direct $X$-$Y$-$\azero$
coupling.\footnote{However, in certain cases we cannot neglect the
  direct $X$-$Y$-$\azero$ coupling.  Actually it is claimed in
  Refs.~\cite{Aceti:2012dj, Aceti:2015zva} that the mixing intensity
  $\xi _{f a}$ is affected by interferences between several diagrams
  of $a_{0}(980)$ and $f_{0}(980)$ productions and hence $\xi _{f a}$
  depends on the reaction, as experimentally observed in the $\eta
  (1405) \to f_{0} (980) \pi ^{0}$ decay~\cite{BESIII:2012aa}.
  Nevertheless, in this study we employ the mixing intensity $\xi _{f
    a}$ in Eq.~\eqref{eq:intensity} since such interferences are
  expected to enhance or decline both the $a_{0} (980)$ and $f_{0}
  (980)$ productions similarly and the mixing intensity $\xi _{f a}$
  will not change so much as long as only intrinsic isospin-violating
  contributions are considered.  On the other hand, if the decaying
  particle exists close to thresholds such as $\bar{K} K^{\ast}$,
  these thresholds could be another source of isospin violation and
  provide a nonnegligible $X$-$Y$-$\azero$ coupling. }  Schematic
diagrams of the $X$ decays to $Y \pi \pi$ and $Y \pi \eta$ are shown
in Fig.~\ref{fig:2}.  In this study we do not take into account
final-state interactions between $Y$ and pseudoscalar mesons by
assuming that decay points of $f_{0}(980)$ and $a_{0}(980)$ are well
isolated from the particle $Y$.  As we will see, the final expression
of the mixing intensity $\xi _{f a}$ does not contain masses nor
widths of the particles $X$ and $Y$.

Since the decay process $X \to Y f_{0}(980) \to Y \pi \pi$ has
three-body final state, the width $\Gamma _{X, f}$ can be
calculated as~\cite{Olive:1900zz}
\begin{align}
\Gamma _{X, f} = & \frac{1}{(2 \pi )^{5} 16 M_{X}^{2}}
\int d M_{\pi \pi} p_{\rm cm} (M_{\pi \pi}) p_{Y} (M_{\pi \pi}) 
\nonumber \\
& \times \int d \Omega \int d \Omega _{Y} | T_{f} |^{2} , 
\label{eq:G-Xf0}
\end{align}
where $M_{X}$ is the mass of the particle $X$, $M_{\pi \pi}$ is the
invariant mass of the $\pi \pi$ system in the final state, and
$\Omega$ and $\Omega _{Y}$ are solid angles for the final $\pi$ in the
$\pi \pi$ rest frame and for the final $Y$ in the $X$ rest frame,
respectively.  Momenta of final-state $\pi$ in the $\pi \pi$ rest
frame, $p_{\rm cm}$, and $Y$ in the $X$ rest frame, $p_{Y}$, are
defined as,
\begin{equation}
p_{\rm cm} ( M )
= \frac{\lambda ^{1/2} (M^{2}, \, m_{\pi}^{2}, \, m_{\pi}^{2})}
{2 M} ,
\end{equation}
\begin{equation}
p_{Y} ( M )
= \frac{\lambda ^{1/2} (M_{X}^{2}, \, M_{Y}^{2}, \, M^{2})}
{2 M_{X}} ,
\end{equation}
respectively, with the particle $Y$ mass $M_{Y}$.  The decay amplitude
$T_{f}$ is expressed as
\begin{equation}
T_{f} = T_{\rm prod} ( M_{\pi \pi} )
P_{f} ( M_{\pi \pi}^{2} )
\bar{g}_{f \pi \pi} ,
\end{equation}
where $T_{\rm prod}$ is the $f_{0}(980)$ production amplitude for the
$X \to Y f_{0}(980)$ process, $P_{f}$ is the $f_{0}(980)$ propagator
with the $a_{0}(980)$-$f_{0}(980)$ mixing given in
Eq.~\eqref{eq:prop_mixing}, and $\bar{g}_{f \pi \pi}$ is the
$f_{0}(980)$-$\pi \pi$ coupling constant in the isospin basis.  Then
we assume that the $f_{0}(980)$ decay width $\Gamma ^{f}$ is small
compared to the energy scales in which the momentum $p_{Y}$ and the
$f_{0}(980)$ production amplitude $T_{\rm prod}$ largely change.  In
this condition, since the $M_{\pi \pi}$ integral is dominated by the
$f_{0}(980)$ mass region due to the $f_{0}(980)$ propagator, we can
approximate the $Y$ momentum $p_{Y}(M_{\pi \pi})$ and the $f_{0}(980)$
production amplitude $T_{\rm prod}(M_{\pi \pi})$ as the values at
$M_{\pi \pi} = M_{f}$, respectively.  Therefore, only the squared
$f_{0}(980)$ propagator $| P_{f} (M_{\pi \pi}^{2}) |^{2}$ and the
momentum $p_{\rm cm}(M_{\pi \pi})$ appear in the $M_{\pi \pi}$
integral in the expression of the decay width~\eqref{eq:G-Xf0}:
\begin{align}
\Gamma _{X, f} = & \frac{1}{( 2 \pi )^{5} 16 M_{X}^{2}} 
p_{Y} ( M_{f} ) \int d \Omega _{Y} | T_{\rm prod} ( M_{f} ) |^{2} 
\nonumber \\
& \times \int d M_{\pi \pi} 4 \pi p_{\rm cm} ( M_{\pi \pi} ) 
| \bar{g}_{f \pi \pi} |^{2} | P_{f} ( M_{\pi \pi}^{2} ) |^{2} ,
\end{align}
where we have performed the integral of the solid angle $\Omega$.
Then by using the relation in Eq.~\eqref{eq:Gamma_abc} and the 
identity $p_{\rm cm}(M) = p_{\pi \pi}(M^{2})$, one can obtain
\begin{equation}
\Gamma _{X, f} = C \times \int d M_{\pi \pi} \, M_{\pi \pi}^{2}
\Gamma _{\pi \pi}^{f} ( M_{\pi \pi}^{2} ) | P_{f} ( M_{\pi \pi}^{2} ) |^{2} ,
\end{equation}
with a constant prefactor $C$:
\begin{equation}
C \equiv \frac{p_{Y}(M_{f})}{16 \pi ^{3} M_{X}^{2}}
\int d \Omega _{Y} | T_{\rm prod} (M_{f}) |^{2} .
\label{eq:C}
\end{equation}

In a similar manner, the width of the decay process $X \to Y
f_{0}(980) \to Y a_{0}(980) \to Y \pi \eta$, $\Gamma _{X, a}$, can be
calculated from
\begin{align}
\Gamma _{X, a} = & \frac{1}{(2 \pi )^{5} 16 M_{X}^{2}}
\int d M_{\pi \eta} p_{\rm cm}^{\prime} (M_{\pi \eta}) p_{Y} (M_{\pi \eta}) 
\nonumber \\
& \times \int d \Omega ^{\prime} \int d \Omega _{Y} | T_{a} |^{2} ,
\end{align}
where $p_{\rm cm}^{\prime}$ is the final-state $\pi$ momentum in the
$\pi \eta$ rest frame,
\begin{equation}
p_{\rm cm}^{\prime} ( M )
= \frac{\lambda ^{1/2} (M^{2}, \, m_{\pi}^{2}, \, m_{\eta}^{2})}
{2 M} ,
\end{equation}
$\Omega ^{\prime}$ is the solid angle for the final $\pi$ in the $\pi
\eta$ rest frame, and $T_{a}$ is the decay amplitude evaluated as
\begin{equation}
T_{a} = T_{\rm prod} ( M_{\pi \eta} )
P_{f \to a} ( M_{\pi \eta}^{2} )
\bar{g}_{a \pi \eta} .
\end{equation}
Here $P_{f \to a}$ is the $f_{0}(980) \to a_{0}(980)$ mixing
propagator given in Eq.~\eqref{eq:prop_mixing} and $\bar{g}_{a \pi
  \eta}$ is the $a_{0}(980)$-$\pi \eta$ coupling constant.  Then, in
order to evaluate the decay width $\Gamma _{X, a}$, we use the fact
that the $a_{0}(980)$-$f_{0}(980)$ mixing takes place particularly at
the $\pi \eta$ invariant mass $M_{\pi \eta} \approx M_{f} \approx
M_{a} \approx 2 m_{K}$.  This is because the $f_{0}(980) \to
a_{0}(980)$ transition is dominated by the difference of the unitarity
cuts for the charged and neutral $K \bar{K}$ thresholds and hence the
mixing amplitude $\Lambda (M_{\pi \eta}^{2})$ shows a narrow peak at
the $K \bar{K}$ thresholds with a width $\sim (m_{K^{0}} +
m_{\bar{K}^{0}}) - (m_{K^{+}} + m_{K^{-}}) \approx 8 \mev$.
Therefore, one can take the values $M_{\pi \eta} \approx M_{f} \approx
M_{a}$ for the amplitude $T_{\rm prod}$ and the momentum $p_{Y}$ in
$\Gamma _{X, a}$.  Moreover, by using the relation in
Eq.~\eqref{eq:Gamma_abc} we can replace the momentum $p_{\rm
  cm}^{\prime} ( M_{\pi \eta} )$ and the squared coupling constant $|
\bar{g}_{a \pi \eta} |^{2}$ with the decay width $\Gamma _{\pi
  \eta}^{a} (M_{\pi \eta}^{2})$ and a kinetic term, which results in
\begin{equation}
\Gamma _{X, a} = C \times \int d M_{\pi \eta} \, M_{\pi \eta}^{2}
\Gamma _{\pi \eta}^{a} ( M_{\pi \eta}^{2} ) | P_{f \to a} ( M_{\pi \eta}^{2} ) |^{2} ,
\end{equation}
where the constant $C$ is same as that in Eq.~\eqref{eq:C}.

As a consequence, we obtain the final expression of the
$a_{0}(980)$-$f_{0}(980)$ mixing intensity $\xi _{f
  a}$~\eqref{eq:intensity} as
\begin{equation}
\xi _{f a} = \frac{\displaystyle
\int d M_{\pi \eta} 
\, M_{\pi \eta}^{2}
\Gamma _{\pi \eta}^{a} ( M_{\pi \eta}^{2} ) | P_{f \to a} ( M_{\pi \eta}^{2} ) |^{2}}
{\displaystyle 
\int d M_{\pi \pi} \, M_{\pi \pi}^{2}
\Gamma _{\pi \pi}^{f} ( M_{\pi \pi}^{2} ) | P_{f} ( M_{\pi \pi}^{2} ) |^{2}} . 
\label{eq:xi_fa}
\end{equation}
The range of the $M_{\pi \eta}$ integral is fixed so as to cover the
$K \bar{K}$ thresholds, say $[0.96 \gev, \, 1.02 \gev ]$.  On the
other hand, we fix the integral range of $M_{\pi \pi}$ so as to take
into account the bump structure coming from the squared propagator
$|P_{f}(M_{\pi \pi}^{2})|^{2}$.  In this formulation, the model
parameters for the mixing intensity are the masses $M_{a}$ and $M_{f}$
in the propagators and the coupling constants $\bar{g}_{a}$,
$\bar{g}_{f}$, $\bar{g}_{a \pi \eta}$, and $\bar{g}_{f \pi \pi}$.  We
note that the final expression of the mixing intensity $\xi _{f a}$
does not contain masses nor widths of the particles $X$ and $Y$, as
one can expect that the $a_{0}(980)$-$f_{0}(980)$ mixing intensity
does not depend on the $f_{0}(980)$ production process.

Finally we mention that one can reproduce the mixing intensity given
in Ref.~\cite{Wu:2008hx} by considering only the integrands in
Eqs.~\eqref{eq:xi_fa} and taking $M_{\pi \pi}^{2} = M_{\pi \eta}^{2} =
s$, which results in
\begin{equation}
\xi _{f a}^{\rm W Z} ( s ) = 
\frac{\Gamma _{\pi \eta}^{a} ( s ) | P_{f \to a} ( s ) |^{2}}
{\Gamma _{\pi \pi}^{f} ( s ) | P_{f} ( s ) |^{2}} .
\label{eq:xi_WZ}
\end{equation}
Actually, in Ref.~\cite{Wu:2008hx} the authors calculated the mixing
intensity $\xi _{f a}^{\rm WZ}$ at the central value of the two $K
\bar{K}$ thresholds, $\xi _{f a}^{\rm WZ} ( (m_{K^{+}} +
m_{K^{0}})^{2} )$.  Here we emphasize that the mixing intensity $\xi
_{f a}^{\rm WZ}$ at the central value of the two $K \bar{K}$
thresholds would be larger than the numerical result from
Eq.~\eqref{eq:xi_fa} with the same parameter set.  This behavior comes
from the fact that in Eq.~\eqref{eq:xi_WZ} we do not perform the
integral of $M_{\pi \pi}$ for the decay $f_{0}(980) \to \pi \pi$.
Namely, while the factor $|P_{f \to a} (s)|^{2}$ has a sharp peak at
the $K \bar{K}$ thresholds due to the $a_{0}(980)$-$f_{0}(980)$
mixing, $|P_{f} (s)|^{2}$ has a relatively broad bump according to the
decay width of $f_{0}(980)$.  Therefore, the numerator of
Eq.~\eqref{eq:xi_WZ} becomes nearly comparable with the denominator
momentarily at the $K \bar{K}$ thresholds, but when they are
integrated the total amount of the numerator becomes only a few or
less than a percent of that of the denominator in
Eq.~\eqref{eq:xi_fa}.  In this study we compare theoretical values of
the mixing intensity with the experimental one, which was obtained as
the ratio of two branching fractions of $J/\psi$, $J/\psi \to \phi
f_{0}(980) \to \phi a_{0}(980) \to \phi \pi \eta$ to $J/\psi \to \phi
f_{0}(980) \to \phi \pi \pi$, so we employ Eq.~\eqref{eq:xi_fa} to
calculate the mixing intensity in the following.

\subsection{Mixing intensity from experimental Flatte parameter sets}
\label{sec:mixing}

\def\arraystretch{1.5}

\begin{table}[!t]
  \caption{Masses and coupling constants of the $a_{0}(980)$ and
    $f_{0}(980)$ resonances in the Flatte
    parametrization~\eqref{eq:Flatte_af} determined from experimental
    data. Here we only show the central values except for the 
    $K \bar{K}$ coupling constant.  Coupling constants are
    given in the isospin basis. }
  \label{tab:1}
  \begin{ruledtabular}
    \begin{tabular*}{8.6cm}{@{\extracolsep{\fill}}lclc}
      \multicolumn{4}{c}{$a_{0}(980)$} \\
      Collaboration & $M_{a}$ [MeV] 
      & $\bar{g}_{a K \bar{K}}$ [GeV] & $\bar{g}_{a \pi \eta}$ [GeV] 
      \\
      \hline
      CLEO~\cite{Adams:2011sq}  &  \phantom{0}998\phantom{.0}  &  
      $3.97 \pm 0.77$  &  4.25 \\
      KLOE~\cite{Ambrosino:2009py}  &  \phantom{0}982.5  &
      $2.84 \pm 0.41$  &  2.46 \\
      CB~\cite{Bugg:2008ig}  &  \phantom{0}987.4  &  
      $2.94 \pm 0.12$  &  2.87 \\
      SND~\cite{Achasov:2000ku}  &  \phantom{0}995\phantom{.0}  &  
      $5.93$ ${}^{+10.54}_{-2.39}$  &  3.11 \\
      E852~\cite{Teige:1996fi}  &  1001\phantom{.0}  &  
      $2.36 \pm 0.13$  &  2.47 \\
      \hline
      \\
      \multicolumn{4}{c}{$f_{0}(980)$}       
      \\
      Collaboration & $M_{f}$ [MeV] 
      & $\bar{g}_{f K \bar{K}}$ [GeV] & $\bar{g}_{f \pi \pi}$ [GeV] 
      \\
      \hline
      CDF~\cite{Aaltonen:2011nk}  &  989.6  &  
      $4.02$ ${}^{+1.01}_{-1.37}$  &  2.65 \\
      KLOE~\cite{Ambrosino:2005wk}  &  977.3  &  
      $2.45 \pm 0.17$  &  1.21 \\
      Belle~\cite{Garmash:2005rv}  &  950\phantom{.0}  &  
      $4.07$ ${}^{+0.76}_{-0.95}$  &  2.28 \\
      BES~\cite{Ablikim:2004wn}  &  965\phantom{.0}  &  
      $5.80$ ${}^{+0.22}_{-0.23}$  &  2.83 \\
      FOCUS~\cite{Link:2004wx}  &  957\phantom{.0}  &  
      $3.39$ ${}^{+0.62}_{-0.76}$  &  2.15 \\
      SND~\cite{Achasov:2000ym}  &  969.8  &  
      $7.88$ ${}^{+1.09}_{-0.86}$  &  3.19 \\
    \end{tabular*}
  \end{ruledtabular}
\end{table}

\begin{table*}[!htb]
  \caption{The $a_{0}(980)$-$f_{0}(980)$ mixing intensity $\xi _{f a}$
    in percentage from the Flatte parameters in Table~\ref{tab:1} with
    the errors for the $K \bar{K}$ coupling constants.  The
    central value of the mixing intensity is shown in bold when it is
    consistent with the experimental value~\eqref{eq:xi_exp}, in italic
    when out of the experimental errors, and with underline when above
    the upper limit~\eqref{eq:xi_exp_UL}.  }
  \label{tab:2}
  \begin{ruledtabular}
    \begin{tabular*}{\textwidth}{@{\extracolsep{\fill}}lc|cccccc}
      & $f_{0} (980)$ & & & & & & \\
      $a_{0} (980)$ & &
      CDF~\cite{Aaltonen:2011nk}  &  
      KLOE~\cite{Ambrosino:2005wk}  &  
      Belle~\cite{Garmash:2005rv}  &  
      BES~\cite{Ablikim:2004wn}  &  
      FOCUS~\cite{Link:2004wx}  &  
      SND~\cite{Achasov:2000ym}  \\
      \hline 
      CLEO~\cite{Adams:2011sq}  & & 
  {\it 0.21} ${} ^{+0.30} _{-0.16}$  &  {\bf 0.53} ${} ^{+0.33} _{-0.23}$  &  {\bf 0.26} ${} ^{+0.30} _{-0.16}$  &  {\bf 0.43} ${} ^{+0.22} _{-0.17}$  &  {\it 0.20} ${} ^{+0.22} _{-0.13}$  &  {\bf 0.73} ${} ^{+0.72} _{-0.38}$  \\
      KLOE~\cite{Ambrosino:2009py} & & 
  {\bf 0.32} ${} ^{+0.40} _{-0.23}$  &  {\bf 0.81} ${} ^{+0.41} _{-0.30}$  &  {\bf 0.38} ${} ^{+0.39} _{-0.23}$  &  {\bf 0.65} ${} ^{+0.26} _{-0.21}$  &  {\bf 0.30} ${} ^{+0.28} _{-0.18}$  &  \underline{\it 1.11} ${} ^{+0.93} _{-0.51}$  \\
      CB~\cite{Bugg:2008ig}  & &
  {\bf 0.26} ${} ^{+0.24} _{-0.17}$  &  {\bf 0.64} ${} ^{+0.18} _{-0.15}$  &  {\bf 0.31} ${} ^{+0.22} _{-0.16}$  &  {\bf 0.52} ${} ^{+0.10} _{-0.09}$  &  {\it 0.24} ${} ^{+0.16} _{-0.12}$  &  {\bf 0.89} ${} ^{+0.50} _{-0.30}$  \\
      SND~\cite{Achasov:2000ku} & & 
  {\bf 0.60} ${} ^{+0.57} _{-0.49}$  &  \underline{\it 1.52} ${} ^{+0.40} _{-0.91}$  &  {\bf 0.70} ${} ^{+0.47} _{-0.52}$  &  \underline{\it 1.22} ${} ^{+0.20} _{-0.69}$  &  {\bf 0.55} ${} ^{+0.35} _{-0.40}$  &  \underline{\it 2.12} ${} ^{+1.00} _{-1.38}$  \\
      E852~\cite{Teige:1996fi} & & 
  {\it 0.19} ${} ^{+0.19} _{-0.13}$  &  {\bf 0.47} ${} ^{+0.14} _{-0.12}$  &  {\it 0.22} ${} ^{+0.17} _{-0.12}$  &  {\bf 0.39} ${} ^{+0.08} _{-0.07}$  &  {\it 0.18} ${} ^{+0.12} _{-0.09}$  &  {\bf 0.66} ${} ^{+0.40} _{-0.23}$  \\
    \end{tabular*}
  \end{ruledtabular}
\end{table*}

\def\arraystretch{1.0}

Since we have formulated the \afmix mixing intensity in the previous
section, we now would like to evaluate the \afmix mixing
intensity~\eqref{eq:xi_fa} with the Flatte
parameters~\eqref{eq:Flatte_af} determined from experimental data.
Actually several collaborations reported the Flatte parameters for
both the $a_{0}(980)$ and $f_{0}(980)$ resonances fitted to the
experimental observations.  In this study we employ parameters by
CLEO~\cite{Adams:2011sq}, KLOE~\cite{Ambrosino:2009py},
CB~\cite{Bugg:2008ig}, SND~\cite{Achasov:2000ku}, and
E852~\cite{Teige:1996fi} for $\azero$, and by
CDF~\cite{Aaltonen:2011nk}, KLOE~\cite{Ambrosino:2005wk},
Belle~\cite{Garmash:2005rv}, BES~\cite{Ablikim:2004wn},
FOCUS~\cite{Link:2004wx}, and SND~\cite{Achasov:2000ym} for $\fzero$.
The parameter sets are listed in Table~\ref{tab:1}.  We note that the
coupling constants in Table~\ref{tab:1} are given in the isospin basis
as in Eq.~\eqref{eq:Gamma_abc}, and especially the $K \bar{K}$
coupling constants in the particle basis, $\bar{g}_{a}$ and
$\bar{g}_{f}$ [see Eq.~\eqref{eq:gKK_PB}], are evaluated with
\begin{equation}
\bar{g}_{a} 
= \frac{1}{\sqrt{2}} \bar{g}_{a K \bar{K}} , 
\quad 
\bar{g}_{f} 
= \frac{1}{\sqrt{2}} \bar{g}_{f K \bar{K}} , 
\end{equation}
where the factor $1 / \sqrt{2}$ translates the coupling constants from
the isospin basis ($\bar{g}_{a K \bar{K}}$ and $\bar{g}_{f K
  \bar{K}}$) into the particle basis.  In this study we take into
account the errors only for the $K \bar{K}$ coupling constants, which
will strongly affect the \afmix mixing intensity, while we take the
central values for other parameters.

The numerical results of the \afmix mixing intensity with all the
combinations of the Flatte parameter sets are given in
Table~\ref{tab:2}.  These values should be compared to the
experimental value~\cite{Ablikim:2010aa}:
\begin{equation}
\xi _{f a} = 0.60 \pm 0.20 _{(\text{stat})} \pm 0.12 _{(\text{sys})}
\pm 0.26 _{(\text{para})} \% , 
\label{eq:xi_exp}
\end{equation}
\begin{equation}
\xi _{f a} |_{\rm upper~limit} = 1.1 \% 
\quad \text{(90\% C.L.)}
\label{eq:xi_exp_UL}
\end{equation}
which was obtained as the ratio of two branching fractions of
$J/\psi$, $J/\psi \to \phi f_{0}(980) \to \phi a_{0}(980) \to \phi \pi
\eta$ to $J/\psi \to \phi f_{0}(980) \to \phi \pi \pi$.  It is
remarkable that two thirds of the combinations of the Flatte parameter
sets reproduce the experimental value with the
errors~\eqref{eq:xi_exp} while only four combinations exceed the
experimental upper limit of the mixing intensity~\eqref{eq:xi_exp_UL}.
We also note that some of the parameter sets tend to lead to small or
large mixing intensity.  For instance, the $\fzero$ parameter set by
FOCUS gives smaller mixing intensity, and the $\azero$ parameter set
by SND gives larger mixing intensity.  Nevertheless, every parameter
set can reproduce the experimental value of the mixing intensity with
a suitable combination.  In this sense we cannot rule out any
parameter set in Table~\ref{tab:1} with the experimental value of the
\afmix mixing intensity.

\section{The compositeness}
\label{sec:comp}

In this study we would like to give a way to extract more information
on the structure of the $\azero$ and $\fzero$ resonances.  For this
purpose, we introduce the compositeness, which corresponds to the
amount of two-body states composing resonances as well as bound
states.  After a brief review of the so-called chiral unitary approach
and compositeness in Sec.~\ref{sec:ChUA}, we calculate the $K \bar{K}$
compositeness of the $\azero$ and $\fzero$ resonances in
Sec.~\ref{sec:KKcomp} by using the Flatte parameter sets.

\subsection{Chiral unitary approach and compositeness}
\label{sec:ChUA}

In this subsection we briefly review the so-called chiral unitary
approach, which provides scattering amplitudes of two pseudoscalar
mesons~\cite{Oller:1997ti, Oller:1997ng, Oller:1998hw, Oller:1998zr}
as well as a pseudoscalar meson and a baryon~\cite{Kaiser:1995eg,
  Oset:1997it, Oller:2000fj, Lutz:2001yb, Jido:2003cb, Hyodo:2011ur}
from the coupled-channel unitarization of the interaction kernel taken
from chiral Lagrangians.  In this approach chiral interactions between
two hadrons dynamically generate hadronic resonances in several
channels from the meson-meson and meson-baryon degrees of freedom with
successful reproductions of experimental observables.  Then, in recent
studies within the chiral unitary approach, structure of the
dynamically generated states is intensively discussed in terms of
compositeness~\cite{Hyodo:2011qc, Aceti:2012dd, Xiao:2012vv,
  Aceti:2013jg, Hyodo:2013nka, Aceti:2014wka, Sekihara:2014}, which
corresponds to the amount of two-body states composing resonances as
well as bound states.  Here we also give the expression of the
compositeness in the chiral unitary approach.

In the chiral unitary approach, we solve the Bethe-Salpeter equation
in an algebraic form so as to obtain a scattering amplitude of two
pseudoscalar mesons
\begin{equation}
T_{i j} ( s ) = V_{i j} ( s ) + \sum _{k}
V_{i k} ( s ) G_{k} ( s ) T_{k j} ( s ) , 
\label{eq:BSeq}
\end{equation}
with channel indices $i$, $j$, and $k$, the Mandelstam variable $s$,
the separable interaction kernel $V$ to be fixed later, and the loop
function $G$ defined as
\begin{align}
& G_{i} ( s ) \equiv 
i \int \frac{d ^{4} q}{(2 \pi)^{4}} 
\frac{1}{q^{2} - m_{i}^{2} + i 0} 
\frac{1}{(P - q)^{2} - m_{i}^{\prime 2} + i 0} 
\nonumber \\
& = \int \frac{d^{3} q}{(2 \pi )^{3}}
\frac{\omega _{i}(\bm{q}) + \omega _{i}^{\prime}(\bm{q})}
{2 \omega _{i}(\bm{q}) \omega _{i}^{\prime}(\bm{q})}
\frac{1}
{s - [\omega _{i}(\bm{q}) + \omega _{i}^{\prime}(\bm{q})]^{2}
+ i 0} ,
\label{eq:Gloop_infty}
\end{align}
where $P^{\mu} = (\sqrt{s}, \, \bm{0})$ and $m_{i}$ and
$m_{i}^{\prime}$ are masses of pseudoscalar mesons in channel $i$.  In
the second line we have performed the $q^{0}$ integral and $\omega
_{i}(\bm{q}) \equiv \sqrt{m_{i}^{2} + \bm{q}^{2}}$ and $\omega
_{i}^{\prime}(\bm{q}) \equiv \sqrt{m_{i}^{\prime 2} + \bm{q}^{2}}$ are
the on-shell energies.

In this construction, a sufficiently strong attractive interaction
with a coupling to an open channel can dynamically generate a
resonance state, which appears as a pole of the scattering amplitude
in the complex lower-half energy plane above the lowest threshold.
The resonance pole is characterized by the pole position and the
residue of the scattering amplitude as
\begin{equation}
T_{i j} ( s ) = \frac{g _{i} g_{j}}{s - s_{\rm pole}} 
+ T_{i j}^{\rm BG} ( s ) , 
\label{eq:amp_pole}
\end{equation}
where $g_{i}$ can be interpreted as the coupling constant of the
resonance to the channel $i$, $\text{Re} \sqrt{s_{\rm pole}}$ ($- 2
\text{Im} \sqrt{s_{\rm pole}}$) corresponds to the mass (width) of the
resonance, and $T_{i j}^{\rm BG}$ is a background term which is
regular at $s \to s_{\rm pole}$.  Then, in Refs.~\cite{Hyodo:2011qc,
  Aceti:2012dd, Xiao:2012vv, Aceti:2013jg, Hyodo:2013nka,
  Aceti:2014wka, Sekihara:2014} the pole position and coupling
constant are further translated into compositeness, which is defined
as the two-body contribution to the normalization of the total wave
function for the resonance.  In our notations of the separable
interaction and the loop function, the $i$-th channel two-body wave
function for resonances generated with the Bethe-Salpeter
equation~\eqref{eq:BSeq} is calculated as~\cite{Sekihara:2014}
\begin{equation}
\tilde{\Psi}_{i} ( \bm{q} ) 
= \frac{g_{i}}
{s_{\rm pole} - [\omega _{i}(\bm{q}) + \omega _{i}^{\prime}(\bm{q})]^{2}} ,
\end{equation}
and the compositeness is obtained as~\cite{Hyodo:2011qc,
  Sekihara:2014}\footnote{For the correct normalization of the
  resonance wave function we do not calculate the absolute value
  squared but the complex number squared of $\tilde{\Psi}(\bm{q})$.
  Moreover, the bra and ket vectors of the resonance state should be
  $\langle \Psi ^{\ast} |$ and $| \Psi \rangle$, respectively, so as
  to obtain the correct normalization, $\langle \Psi ^{\ast} | \Psi
  \rangle = 1$ (see Refs.~\cite{Hyodo:2013nka, Sekihara:2014} for
  details).}
\begin{align}
X_{i} \equiv & \int 
\frac{d^{3} q}{( 2 \pi )^{3}}
\frac{\omega _{i}(\bm{q}) + \omega _{i}^{\prime}(\bm{q})}
{2 \omega _{i}(\bm{q}) \omega _{i}^{\prime}(\bm{q})} 
\left [ \tilde{\Psi}_{i} ( \bm{q} ) \right ]^{2}
\notag \\ 
= & - g_{i}^{2} \frac{d G_{i}}{d s} ( s = s_{\rm pole} ) ,
\label{eq:Xi}
\end{align}
where the normalization factor $[\omega _{i}(\bm{q}) + \omega
_{i}^{\prime}(\bm{q})] / [2 \omega _{i}(\bm{q}) \omega
_{i}^{\prime}(\bm{q})]$ guarantees the Lorentz invariance of the
integral and in the last line the integral is transformed into the
derivative of the loop function $G_{i}$~\eqref{eq:Gloop_infty}.  We
note that the compositeness is not an observable and hence is a model
dependent quantity.  We also note that the derivative of the loop
function does not diverge for meson-meson states in contrast to the
loop function itself, but one has to treat consistently the loop
function and its derivative, i.e., one has to use the same
regularization for both the loop function and its derivative.  On the
other hand, in order to measure the fraction of the bare state
contribution rather than the two-body state involved, we introduce the
elementariness $Z$, which corresponds to the field renormalization
constant intensively discussed in the 1960s~\cite{Salam:1962ap,
  Weinberg:1962hj, Ezawa:1963zz, Weinberg:1965zz}.  The elementariness
measures all contributions which cannot be responsible for the
hadronic two-body component involved.  For instance, compact $q
\bar{q}$ and $q q \bar{q} \bar{q}$ states contribute to the
elementariness.  The expression of the elementariness in our notations
is obtained in Ref.~\cite{Sekihara:2014} as
\begin{equation}
Z = - \sum _{i,j} g_{j} g_{i} \left [ G_{i} \frac{d V_{i j}}{d s} G_{j} 
\right ] _{s = s_{\rm pole}} . 
\label{eq:Z}
\end{equation}
We note that in general both the compositeness $X_{i}$ and
elementariness $Z$ take complex values for a resonance state
and hence one cannot interpret the compositeness (elementariness) as
the probability to observe a two-body (bare-state) component inside
the resonance.  However, a striking property of the compositeness and
elementariness is that sum of them coincides with the normalization of
the total wave function for the resonance $| \Psi \rangle$ and is exactly
unity~\cite{Sekihara:2014}:
\begin{align}
  \langle \Psi ^{\ast} | \Psi \rangle = & \sum _{i} X_{i} + Z 
  \notag \\
  = & - \sum _{i,j} g_{j} g_{i} 
  \left [ \delta _{ij} \frac{d G_{i}}{d s} 
    + G_{i} \frac{d V_{i j}}{d s} G_{j} 
  \right ] _{s = s_{\rm pole}}
  = 1 ,
\label{eq:sum-rule}
\end{align}
where the condition of the correct normalization as unity is
guaranteed by a generalized Ward identity proved in
Ref.~\cite{Sekihara:2010uz}.  Based on this normalization, we propose
to interpret the compositeness $X_{i}$ and elementariness $Z$ for a
certain class of resonances on the basis of the similarity to the
bound state case.  Namely, if the compositeness $X_{1}$ approaches
unity with a small imaginary part while $X_{i}$ ($i \ne 1$) and $Z$
negligibly contribute to the normalization~\eqref{eq:sum-rule}, the
system can be interpreted to be dominated by the two-body component in
channel $1$, since the resonance wave function is considered to be
similar to that of the bound state dominated by the channel $1$.  In
this sense, $|X_{1}| \sim 1$ is a necessary condition for the
molecular picture in channel $1$.  In another case, if $| X_{i} |$ is
much smaller than unity, the system contains negligible $i$-channel
two-body component.


In order to examine the chiral unitary approach and compositeness, let
us consider scatterings of $s$-wave two pseudoscalar mesons which
couples to the $a_{0}(980)$ and $f_{0}(980)$ resonances.  We here
assume the isospin symmetry and introduce five channels labeled by the
indices $i=1, \dots , 5$ in the order $K^{+}K^{-}$, $K^{0}
\bar{K}^{0}$, $\pi ^{+} \pi ^{-}$, $\pi ^{0} \pi ^{0}$, and $\pi ^{0}
\eta$.  The interaction kernel $V_{i j} = V_{j i}$ is taken from the
leading-order chiral Lagrangian as
\begin{equation}
\begin{split}
& V_{11} = 2 V_{12} = 2 V_{13} = 2 \sqrt{2} V_{14} 
\\
& = V_{22} = 2 V_{23} = 2 \sqrt{2} V_{24} = V_{33} 
= - \frac{s}{2 f^{2}} , 
\\
& V_{15} = - V_{25} = - \frac{3 s - 4 m_{K}^{2}}{4 \sqrt{3} f^{2}} ,
\label{eq:Vint} 
\\
& V_{34} = - \frac{s - m_{\pi}^{2}}{\sqrt{2} f^{2}} , 
\\
& V_{35} = V_{45} = 0 , 
\\
& V_{44} = \frac{3}{2} V_{55} = - \frac{m_{\pi}^{2}}{2 f^{2}} , 
\end{split}
\end{equation}
with the pion decay constant $f$.  We note that we have multiplied the
interaction kernel~\eqref{eq:Vint} in the case of $\pi ^{0} \pi ^{0}$
states by $1/\sqrt{2}$ compared to the expression given in, {\it
  e.g.}, Ref.~\cite{Hanhart:2007bd}, thus a na\"{i}ve unitarization in
Eq.~\eqref{eq:BSeq} with the interaction kernel~\eqref{eq:Vint} can
give a correct normalization for intermediate states of identical
particles.  On the other hand, for the loop function we employ a
three-dimensional cut-off as
\allowdisplaybreaks[0]
\begin{align}
& G_{i} ( s ; \, q_{\rm max} ) 
\nonumber \\
& = \int \frac{d^{3} q}{(2 \pi )^{3}}
\frac{\omega _{i}(\bm{q}) + \omega _{i}^{\prime}(\bm{q})}
{2 \omega _{i}(\bm{q}) \omega _{i}^{\prime}(\bm{q})}
\frac{\theta ( q_{\rm max} - |\bm{q}| )}
{s - [\omega _{i}(\bm{q}) + \omega _{i}^{\prime}(\bm{q})]^{2}
+ i 0} ,
\label{eq:Gloop}
\end{align}
where $\theta (x)$ is the Heaviside step function.

\begin{table}
  \caption{Properties of the $a_{0}(980)$ and $f_{0}(980)$ resonances
    in the chiral unitary approach.  For later convenience we show 
    the values in the particle basis. }
  \label{tab:3}
  \begin{ruledtabular}
    \begin{tabular*}{8.6cm}{@{\extracolsep{\fill}}ccc}
      & $a_{0}(980)$ & $f_{0}(980)$ \\
      \hline
      $\sqrt{s_{\rm pole}}$ & $977 - 56 i \mev$ & $983 - 21 i \mev$ \\
      $g_{K^{+} K^{-}}$ 
      & $3.31  +0.28 i \gev$
      & $2.97  +0.89 i \gev$ \\
      $g_{K^{0} \bar{K}^{0}}$ 
      & $-3.31  -0.28 i \gev \phph$
      & $2.97  +0.89 i \gev$ \\
      $g_{\pi ^{+} \pi ^{-}}$ 
      & --- 
      & $-0.29  +1.28 i \gev \phph$ \\
      $g_{\pi ^{0} \pi ^{0}}$ 
      & ---
      & $-0.20  +0.90 i \gev \phph$ \\
      $g_{\pi ^{0} \eta}$ 
      & $2.99  -0.85 i \gev$
      & --- \\
      $X_{K^{+} K^{-}}$ 
      & $0.17 -0.15 i$
      & $0.35 -0.05 i$ \\
      $X_{K^{0} \bar{K}^{0}}$ 
      & $0.17 -0.15 i$
      & $0.35 -0.05 i$ \\
      $X_{\pi ^{+} \pi ^{-}}$ 
      & --- 
      & $0.01 + 0.01 i$ \\
      $X_{\pi ^{0} \pi ^{0}}$ 
      & --- 
      & $0.01 + 0.00 i$ \\
      $X_{\pi ^{0} \eta}$ 
      & $-0.07 + 0.12 i \phph$ 
      & --- \\
      $Z$ 
      & $0.73 + 0.18 i$ 
      & $0.28 + 0.10 i$ \\
    \end{tabular*}
  \end{ruledtabular}
\end{table}

Now we solve the Bethe-Salpeter equation~\eqref{eq:BSeq} with the
isospin symmetric masses and parameters $q_{\rm max}=1.075 \gev$ and
$f=93.0 \mev$, which are chosen so as to generate two poles which
correspond to the $a_{0} (980)$ and $f_{0}(980)$ resonances,
respectively, in the complex $s$ plane of the scattering amplitude.
The pole positions, coupling constants, compositeness, and
elementariness of the two resonances are listed in Table~\ref{tab:3}.
We note that for the evaluations of the compositeness and
elementariness we use the same sharp cut-off $q_{\rm max}$ for the
loop function and its derivative.  As a result, the $K \bar{K}$
compositeness for the resonance, $X_{K \bar{K}}$, can be obtained by
summing up the $i=1$ ($K^{+} K^{-}$) and $2$ ($K^{0} \bar{K}^{0}$)
contributions as
\begin{equation}
X_{K \bar{K}} \equiv - \sum _{i=1}^{2}
g_{i}^{2} \frac{d G_{i}}{d s} ( s = s_{\rm pole}; \, q_{\rm max} ) ,
\label{eq:XKKbar}
\end{equation}
which results in $0.34 - 0.30 i$ for $a_{0}(980)$ and $0.70 - 0.11 i$
for $f_{0}(980)$.  Since the real part dominates the sum
rule~\eqref{eq:sum-rule} while the imaginary part is negligible, the
$K \bar{K}$ compositeness for $f_{0}(980)$ indicates a large $K
\bar{K}$ component inside it, on the basis of the similarity to the
bound state case.  This finding is consistent with a model-independent
analysis in Ref.~\cite{Baru:2003qq}.  The $K \bar{K}$ compositeness
for $a_{0}(980)$ implies a nonnegligible $K \bar{K}$ component inside
it, but we cannot clearly conclude the structure due to its large
imaginary part.  The difference of the structure of $a_{0}(980)$ and
$f_{0}(980)$ may originate from the fact that the strength of the
leading-order $K \bar{K} (I=0)$ interaction from chiral perturbation
theory [see Eq.~\eqref{eq:Vint}] is three times larger than that with
$I=1$:
\begin{equation}
V_{K \bar{K} (I=0)} = - \frac{3 s}{4 f^{2}} , 
\quad 
V_{K \bar{K} (I=1)} = - \frac{s}{4 f^{2}} . 
\end{equation}
In addition, the absolute values of the $\pi \eta$ and $\pi \pi$
compositeness are much smaller than unity, so both the $\pi \eta$
component inside $a_{0}(980)$ and the $\pi \pi$ component inside
$f_{0}(980)$ are negligible.

At the end of this subsection we emphasize that, although the
compositeness is not an observable, we can evaluate it from
experimental observables via appropriate models. In the following we
employ the expression in Eq.~\eqref{eq:XKKbar} for the $K \bar{K}$
compositeness of the $a_{0}(980)$ and $f_{0}(980)$ resonances.  We
will take the cut-off $q_{\rm max} \to \infty$ for the derivative of
the loop function $G_{i}$; the use of finite cut-off $q_{\rm max} \sim
1 \gev$ will give only several percent change of the value of the
compositeness.

\subsection{$\bm{K \bar{K}}$ compositeness from experimental Flatte
  parameter sets}
\label{sec:KKcomp}

As we have seen in the previous subsection, the $K \bar{K}$
compositeness of the $\azero$ and $\fzero$ mesons~\eqref{eq:XKKbar}
can be determined with their pole positions and residues of the
scattering amplitudes.  Here we adopt the Flatte parametrization
without the mixing in Eq.~\eqref{eq:Flatte_af}, and we calculate the
pole positions, residues, and compositeness for the scalar mesons from
experimentally fitted parameters in Table~\ref{tab:1}.  Namely, the
propagator of the scalar meson $A$ ($A = a$, $f$), $1 / D_{A}(s)$,
brings a pole in the scattering amplitude of $T_{i j}^{A}$:
\begin{equation}
T_{i j}^{A} = \frac{\bar{g}_{A i} \bar{g}_{A j}}{D_{A} ( s )} 
+ (\text{regular at }s = s_{A}) , 
\end{equation}
where $s_{A}$ is the pole position of $1 / D_{A}(s)$ and $\bar{g}_{A
  i} \bar{g}_{A j}$ is multiplied so as to describe the scattering $i$
to $j$.  Therefore, compared to Eq.~\eqref{eq:amp_pole} the residue at
the resonance pole of the scattering amplitude $T^{A}$ can be
evaluated as
\begin{equation}
g_{A} ^{2} = \bar{g}_{A}^{2} R_{A} , 
\label{eq:bargA}
\end{equation}
\begin{equation}
R_{A} \equiv \text{Res} [ 1 / D_{A} ( s ); \, s_{A} ] 
= \lim _{s \to s_{A}} \frac{s - s_{A}}{D_{A} ( s )} .
\label{eq:ResA}
\end{equation}
As a consequence, the $K \bar{K}$ compositeness for the resonance $A$,
$X_{A}$, can be obtained by summing up the $i=1$ ($K^{+} K^{-}$) and
$2$ ($K^{0} \bar{K}^{0}$) contributions as
\begin{align}
X_{A} = & - g_{A}^{2} \sum _{i=1}^{2}
\frac{d G_{i}}{d s} ( s = s_{A} ; \, \infty ) 
\nonumber \\
= & - \bar{g}_{A}^{2} R_{A} \sum _{i=1}^{2}
\frac{d G_{i}}{d s} ( s = s_{A} ; \, \infty ) .
\end{align}
This is the formula to calculate the $K \bar{K}$ compositeness of the
scalar mesons from the Flatte parametrization without the mixing.  In
a similar manner we can calculate the $\pi \pi$ and $\pi \eta$
compositeness for the scalar mesons.

\begin{table}[!t]
  \caption{Pole positions and compositeness from the Flatte
    parameters given in Table~\ref{tab:1}. Here we only show the central
    values.  Compositeness is given in the isospin basis.}
  \label{tab:4}
  \begin{ruledtabular}
    \begin{tabular*}{8.6cm}{@{\extracolsep{\fill}}lccc}
      \multicolumn{4}{c}{$a_{0}(980)$} \\
      Collaboration & $\sqrt{s_{a}}$ [MeV] & $X_{a}$ & $X_{\pi \eta}$
      \\
      \hline
      CLEO~\cite{Adams:2011sq}  &  
      $ 1022  -70 i$  &  $0.09  -0.22 i$  &  $-0.16  +0.05 i$ 
      \\
      KLOE~\cite{Ambrosino:2009py} & 
      $\phantom{0}994 - 26 i$ & $0.15 -0.17 i$ & $-0.05 +0.03 i$ 
      \\
      CB~\cite{Bugg:2008ig} & 
      $1000 - 36 i$ & $0.12 -0.17 i$ & $-0.07 +0.03 i$ 
      \\
      SND~\cite{Achasov:2000ku} & 
      $1005 - \phantom{0}5 i$ & $0.68 -0.51 i$ & $-0.05 -0.01 i$ 
      \\
      E852~\cite{Teige:1996fi} &
      $ 1007  -28 i$  &  $0.07  -0.15 i$  &  $-0.06  +0.03 i$ 
      \\
      \hline
      \\
      \multicolumn{4}{c}{$f_{0}(980)$} 
      \\
      Collaboration & $\sqrt{s_{f}}$ [MeV] & $X_{f}$ & $X_{\pi \pi}$ 
      \\
      \hline
      CDF~\cite{Aaltonen:2011nk}  &  
      $ 1010  -30 i$  &  $0.21  -0.30 i$  &  $-0.03  -0.01 i$ \\
      KLOE~\cite{Ambrosino:2005wk} & 
      $\phantom{0}985 - 10 i$ & $0.21 -0.12 i$ & $-0.01 -0.00 i$ 
      \\
      Belle~\cite{Garmash:2005rv}  &  
      $\phantom{0}983  -27 i$  &  $0.29  -0.17 i$  &  $-0.02  -0.00 i$ \\
      BES~\cite{Ablikim:2004wn} & 
      $1000 - 18 i$ & $0.50 -0.34 i$ & $-0.02 -0.01 i$ 
      \\
      FOCUS~\cite{Link:2004wx}  &  
      $\phantom{0}981  -28 i$  &  $0.21  -0.14 i$  &  $-0.02  -0.00 i$ \\
      SND~\cite{Achasov:2000ym} & 
      $1001 - \phantom{0}7 i$ & $0.80 -0.31 i$ & $-0.01 -0.01 i$ 
      \\
    \end{tabular*}
  \end{ruledtabular}
\end{table}%

\begin{figure}[!t]
  \centering
  \Psfig{8.6cm}{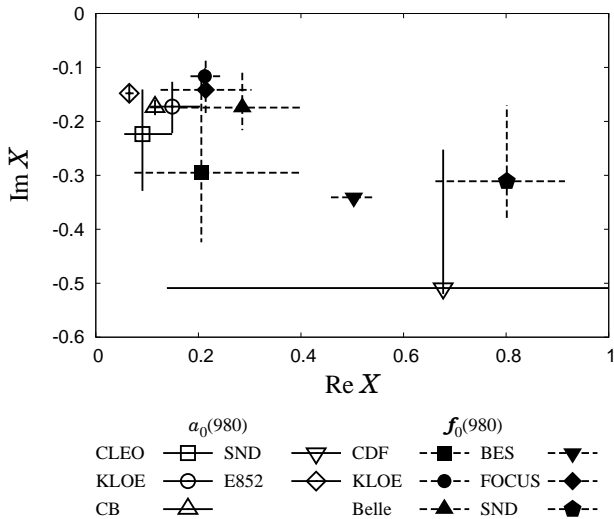}
  \caption{The $K \bar{K}$ compositeness of the $\azero$ and $\fzero$
    resonances from the the Flatte parameters in Table~\ref{tab:1}
    with the errors for the $K \bar{K}$ coupling
    constants.  Open (filled) symbols with solid (dashed) lines
    represent the $K \bar{K}$ compositeness for $\azero$ [$\fzero$]. }
\label{fig:KKcomp_Flatte}
\end{figure}

Now we calculate the pole positions and compositeness of the $\azero$
and $\fzero$ scalar mesons from the parameter sets in
Table~\ref{tab:1}.  The numerical results are listed in
Table~\ref{tab:4} and plotted in Fig.~\ref{fig:KKcomp_Flatte}.  In the
table we show only the central values, while we take into account the
errors for the $K \bar{K}$ coupling constants in the figure.  All of
the pole positions in Table~\ref{tab:4} exist in the physical Riemann
sheet of the $K \bar{K}$ channel.

As one can see from Table~\ref{tab:4} and
Fig.~\ref{fig:KKcomp_Flatte}, we obtain the complex compositeness in
every parameter sets since $\azero$ and $\fzero$ are resonance states.
For the $K \bar{K}$ compositeness of $a_{0}(980)$, the parameters do
not give large absolute value of the $K \bar{K}$ compositeness
comparable to unity except for the SND parameter, which however has a
large error bar as seen in Fig.~\ref{fig:KKcomp_Flatte}.  The absolute
value $| X_{a} | \sim 0.2$ in other parameter sets could imply a small
but nonnegligible $K \bar{K}$ component inside $\azero$, but at
present we do not clearly interpret the $K \bar{K}$ compositeness for
$\azero$.  On the other hand, for $f_{0}(980)$ two of the parameter
sets (BES and SND) imply large absolute value of the $K \bar{K}$
component with, say $| X_{f} |>0.6$.  Especially it is worth noting
that in the BES analysis the authors fitted the $K \bar{K}$ spectrum
as well as the $\pi \pi$ spectrum, which leads to a small error coming
from the $K \bar{K}$ coupling constant.  We note that the tendency for
$\fzero$ supports the result in the chiral unitary approach (see
Table~\ref{tab:3} and also Ref.~\cite{Sekihara:2014,
  Sekihara:2012xp}).

The tendency for $\fzero$ is similar to the findings in
Ref.~\cite{Baru:2003qq}, which suggested that $f_{0}(980)$ should be a
$K \bar{K}$ molecular state to a large degree.  However, although both
results in the present study and in Ref.~\cite{Baru:2003qq} are
obtained with the experimental Flatte parameters and similar
formulations, a big difference is that in Ref.~\cite{Baru:2003qq} they
defined ``compositeness'' in terms of the spectral density as a real
value even for the $\azero$ and $\fzero$ resonances.  This treatment
should be valid effectively only when the resonance has a narrow width
and the imaginary part of the compositeness is enough small (see also
Ref.~\cite{Hyodo:2013nka}).  Otherwise, the correct
normalization~\eqref{eq:sum-rule} will be lost.  However, our result
gives a nonnegligible imaginary part of the compositeness from the
Flatte parameters, which means that the treatment in
Ref.~\cite{Baru:2003qq} should be reexamined.  In other words, the
value calculated in Ref.~\cite{Baru:2003qq} should not be compared
with unity, since the correct normalization of the resonance wave
function should be lost.

We also note that all of the absolute values of the $\pi \eta$ and
$\pi \pi$ compositeness for the $\azero$ and $\fzero$ resonances,
respectively, are small compared to unity.  This strongly indicates
that the $\azero$ and $\fzero$ resonances are not the $\pi \eta$ and
$\pi \pi$ molecular states, respectively.

\section{Constraint on the $\bm{K \bar{K}}$ compositeness from the
  $\bm{a_{0}(980)}$-$\bm{f_{0}(980)}$ mixing intensity}
\label{sec:relation}

\subsection{Relation between the $\bm{a_{0}(980)}$-$\bm{f_{0}(980)}$
  mixing intensity and the $\bm{K \bar{K}}$ compositeness from
  experimental Flatte parameter sets}

Now that we have formulated both the $a_{0}(980)$-$f_{0}(980)$ mixing
intensity $\xi _{f a}$ and the compositeness $X$, we would like to
investigate a relation between them for the $a_{0}(980)$ and
$f_{0}(980)$ resonances.  To this end we first mention that the mixing
amplitude $\Lambda (s)$~\eqref{eq:mix_amp} is proportional to the
product of the coupling constants, $\bar{g}_{a} \bar{g}_{f}$.  This
indicates that, when the product of the coupling constants
$\bar{g}_{a} \bar{g}_{f}$ and hence the mixing amplitude $\Lambda (s)$
are sufficiently small, the mixing intensity behaves in power of
\begin{equation}
\xi _{f a} \sim | \Lambda |^{2} \sim | \bar{g}_{a} \bar{g}_{f} |^{2} 
\propto |X_{a} X_{f}| .
\label{eq:xi_behave}
\end{equation}
Therefore, we expect that the mixing intensity $\xi _{f a}$ is
proportional to the absolute value of the product of the $K \bar{K}$
compositeness of $a_{0}(980)$ and $f_{0}(980)$, $| X_{a} X_{f} |$, for
small $K \bar{K}$ coupling constants.

\begin{figure}[!t]
  \centering
  \Psfig{8.6cm}{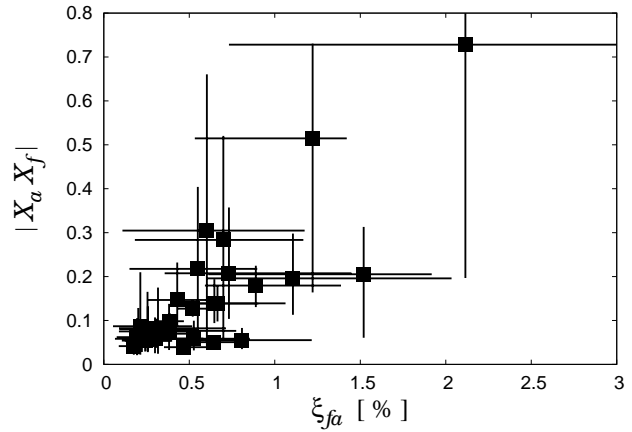}
  \caption{Scatter plot of the absolute value of the product of the $K
    \bar{K}$ compositeness for the $a_{0}(980)$ and $f_{0}(980)$
    resonances, $|X_{a} X_{f}|$, with respect to the
    $a_{0}(980)$-$f_{0}(980)$ mixing intensity $\xi _{f a}$.  Data
    points are obtained from the Flatte parameters given in
    Table~\ref{tab:1} with the errors for the $K \bar{K}$ coupling
    constants.}
\label{fig:XaXf_xi_Flatte}
\end{figure}

In order to examine the behavior in Eq.~\eqref{eq:xi_behave}, we plot
in Fig.~\ref{fig:XaXf_xi_Flatte} the absolute value of the product of
the $K \bar{K}$ compositeness for the $a_{0}(980)$ and $f_{0}(980)$
resonances, $|X_{a} X_{f}|$, with respect to the
$a_{0}(980)$-$f_{0}(980)$ mixing intensity $\xi _{f a}$ by using the
Flatte parameters in Table~\ref{tab:1}.  As one can see from
Fig.~\ref{fig:XaXf_xi_Flatte}, although a clear proportional
connection $\xi _{f a} \propto |X_{a} X_{f}|$ is not observed, there
is a tendency that the product $|X_{a} X_{f}|$ increases as the mixing
intensity $\xi _{f a}$ increases.  From the experimental upper limit
of the mixing intensity~\eqref{eq:xi_exp_UL}, we expect that the
product of the compositeness has a upper bound as $|X_{a} X_{f}|
\lesssim 0.4$.  This upper bound implies that the $a_{0}(980)$ and
$f_{0}(980)$ resonances cannot be simultaneously $K \bar{K}$ molecular
states, since the condition $|X_{a}| \sim |X_{f}| \sim 1$ cannot
satisfy $|X_{a} X_{f}| \lesssim 0.4$.

We note that the above discussion is based on the Flatte parameters in
Table~\ref{tab:1}.  Therefore, in order to confirm this relation
between the mixing intensity and the absolute value of the $K \bar{K}$
compositeness for $a_{0}(980)$ and $f_{0}(980)$, we have to calculate
them with more general parameter sets, especially for the $\azero$-
and $\fzero$-$K \bar{K}$ coupling constants, which are responsible for
both the mixing intensity and their $K \bar{K}$ compositeness.  This
is the task in the next subsection.

\subsection{Confirmation of the relation}

\subsubsection{Strategy}

Now we construct a relation between the $a_{0}(980)$-$f_{0}(980)$
mixing intensity $\xi _{f a}$ and the compositeness $X$ for the
$a_{0}(980)$ and $f_{0}(980)$ resonances.  Our strategy is summarized
as follows.  First, we fix four parameters $M_{a}$, $M_{f}$,
$\bar{g}_{a \pi \eta}$, and $\bar{g}_{f \pi \pi}$ in some appropriate
approaches.  Then we generate the $a_{0}(980)$- and $f_{0}(980)$-$K
\bar{K}$ coupling constants, $\bar{g}_{a}$ and $\bar{g}_{f}$,
respectively, which are responsible for both the mixing intensity $\xi
_{f a}$~\eqref{eq:xi_fa} and their $K \bar{K}$ compositeness, to
evaluate simultaneously the mixing intensity and their $K \bar{K}$
compositeness.  With this approach we can give a more general
constraint on the $K \bar{K}$ structure of $a_{0} (980)$ and $f_{0}
(980)$ regardless of the details of the $a_{0} (980)$- and $f_{0}
(980)$-$K \bar{K}$ coupling constants.

For the $K \bar{K}$ compositeness we employ the model in
Sec.~\ref{sec:comp}, and evaluate it with the following expression:
\begin{equation}
X_{A} = - g_{A}^{2} \sum _{i=1}^{2}
\frac{d G_{i}}{d s} ( s = s_{A} ; \, \infty ) ,
\quad 
A = a, \, f .
\end{equation}
In order that we can calculate the $K \bar{K}$ compositeness with not
only small but also large mixing amplitude, the pole position $s_{A}$
for the resonance $A$ is extracted as that of the propagator
$P_{A}(s)$ rather than the propagator without mixing, $1 / D_{A}(s)$.
We note that the coupling constant $g_{A}$ in the expression of the
compositeness should be evaluated as a residue of the resonance pole
position [see Eq.~\eqref{eq:amp_pole}], and hence it differs from
$\bar{g}_{A}$ by the residue of the propagator $P_{A}(s)$.  Namely, in
a similar manner to the discussion in Sec~\ref{sec:KKcomp}, taking
into account the residue of the propagator $P_{A}(s)$ as
\begin{equation}
g_{A} ^{2} = \bar{g}_{A}^{2} R_{A}^{\prime} , 
\quad 
R_{A}^{\prime} \equiv \lim _{s \to s_{A}} (s - s_{A}) P_{A} ( s ) ,
\end{equation}
we can calculate the compositeness as
\begin{equation}
X_{A} = - \bar{g}_{A}^{2} R_{A}^{\prime} \sum _{i=1}^{2}
\frac{d G_{i}}{d s} ( s = s_{A} ; \, \infty ) .
\label{eq:XA_mix}
\end{equation}

Although the compositeness $X_{A}$ is in general complex for resonance
states, in this study we use Eq.~\eqref{eq:XA_mix} so as to evaluate
the absolute value of the compositeness $|X_{A}|$ from the coupling
constant $\bar{g}_{A}$.  Therefore, in our strategy we will obtain a
relation between the mixing intensity and the absolute value of the $K
\bar{K}$ compositeness for $a_{0}(980)$ and $f_{0}(980)$.  We
emphasize that the absolute value of the compositeness cannot be
interpreted as a probability to find a two-body molecular state but it
will be an important piece of information on the structure of the
scalar mesons when compared with unity.  For instance, $|X_{A}| \sim
1$ is a necessary condition for the $K \bar{K}$ molecular picture of
the meson $A$, while $|X_{A}| \ll 1$ indicates that the meson $A$ has
a negligible $K \bar{K}$ molecular component.

\subsubsection{Relation between the \afmix mixing intensity and the $K
  \bar{K}$ compositeness}

We now fix the four parameters as rough averages of the Flatte
parameters listed in Table~\ref{tab:1}:
\begin{equation}
\begin{split}
& M_{a} = 990 \mev , 
\quad 
\bar{g}_{a \pi \eta} = 3.0 \gev ,
\\
& M_{f} = 970 \mev , 
\quad 
\bar{g}_{f \pi \pi} = 2.4 \gev ,
\label{eq:af_param}
\end{split}
\end{equation}
to construct a relation between the $K \bar{K}$ compositeness of the
$a_{0}(980)$ and $f_{0}(980)$ resonances and their mixing intensity
$\xi _{f a}$.  Since we expect that the mixing intensity behaves as in
Eq.~\eqref{eq:xi_behave} for small $K \bar{K}$ coupling constants, we
investigate a relation between the mixing intensity $\xi _{f a}$ and
the absolute value of the product of the $K \bar{K}$ compositeness of
$a_{0}(980)$ and $f_{0}(980)$, $| X_{a} X_{f} |$.  Here we employ the
Monte-Carlo method and generate values of the coupling constants
$\bar{g}_{a}$ and $\bar{g}_{f}$ independently from random numbers.  In
the Monte-Carlo method we take the range of the $K \bar{K}$ coupling
constants as $[0 \gev , \, 6 \gev ]$ both for $\bar{g}_{a}$ and
$\bar{g}_{f}$, which covers the values of the coupling constants
$\bar{g}_{a K \bar{K}}$ ($= \sqrt{2} \bar{g}_{a}$) and $g_{f K
  \bar{K}}$ ($= \sqrt{2} \bar{g}_{f}$) listed in Table~\ref{tab:1}.

\begin{figure}[!t]
  \centering
  \Psfig{8.6cm}{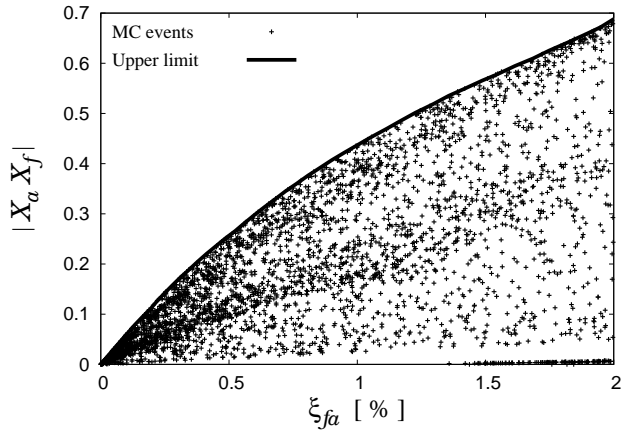}
  \caption{Scatter plot of the absolute value of the product of the $K
    \bar{K}$ compositeness for the $a_{0}(980)$ and $f_{0}(980)$
    resonances, $|X_{a} X_{f}|$, with respect to the
    $a_{0}(980)$-$f_{0}(980)$ mixing intensity $\xi _{f a}$.  In the
    plot, number of data in the Monte-Carlo method (MC events) amounts
    to $\sim 4 \times 10^{3}$.  We also show an upper limit of $|X_{a}
    X_{f}|$ for each value of $\xi _{f a}$ by the solid line.}
\label{fig:XaXf_xi}
\end{figure}

In Fig.~\ref{fig:XaXf_xi} we show a scatter plot for the absolute
value of the product of the $K \bar{K}$ compositeness of $a_{0}(980)$
and $f_{0}(980)$, $| X_{a} X_{f} |$, with respect to the mixing
intensity $\xi _{f a}$, with various values of the coupling constants
$\bar{g}_{a}$ and $\bar{g}_{f}$ from random numbers.  As one can see
from Fig.~\ref{fig:XaXf_xi}, although a proportional connection $\xi
_{f a} \propto |X_{a} X_{f}|$ is not observed, no Monte-Carlo data
point exists in the upper-left region of the plot, which implies that
there is an upper limit of allowed $|X_{a} X_{f}|$ for each value of
$\xi _{f a}$.  In fact we can check the existence of this upper limit
by sweeping the values of $\bar{g}_{a}$ and $\bar{g}_{f}$
independently in appropriate ranges, say $[0 \gev , \, 6 \gev ]$, and
the result of the upper limit of $|X_{a} X_{f}|$ for each $\xi _{f a}$
is plotted as a solid line in Fig.~\ref{fig:XaXf_xi}.  We note that
the upper limit of the allowed $|X_{a} X_{f}|$ behaves like $|X_{a}
X_{f}|_{\rm upper~limit} \sim \xi _{f a}$.  This result means that, if
we observe smaller value of the mixing intensity $\xi _{f a}$, the
product of the compositeness $|X_{a} X_{f}|$ also becomes smaller.

We then discuss the compositeness of the scalar mesons with the
experimental value of the mixing intensity~\eqref{eq:xi_exp} and
\eqref{eq:xi_exp_UL}, which was obtained as the ratio of two branching
fractions of $J/\psi$, $J/\psi \to \phi f_{0}(980) \to \phi a_{0}(980)
\to \phi \pi \eta$ to $J/\psi \to \phi f_{0}(980) \to \phi \pi \pi$.
This experimental value, together with the upper limit of $|X_{a}
X_{f}|$ for each value of the mixing intensity $\xi _{f a}$ shown in
Fig.~\ref{fig:XaXf_xi}, can constrain the structure of $a_{0}(980)$
and $f_{0}(980)$ through their compositeness.  For instance, with
Fig.~\ref{fig:XaXf_xi}, $\xi _{f a} |_{\rm upper~limit} = 1.1 \%$
gives a constraint $|X_{a} X_{f}| < 0.47$.  With this constraint, we
confirm that both the $a_{0}(980)$ and $f_{0}(980)$ resonances are
simultaneously $K \bar{K}$ molecular states is questionable, since the
condition $|X_{a}| \sim |X_{f}| \sim 1$ is out of the allowed region
$|X_{a} X_{f}| < 0.47$.

\begin{figure}[!t]
  \centering
  \Psfig{8.6cm}{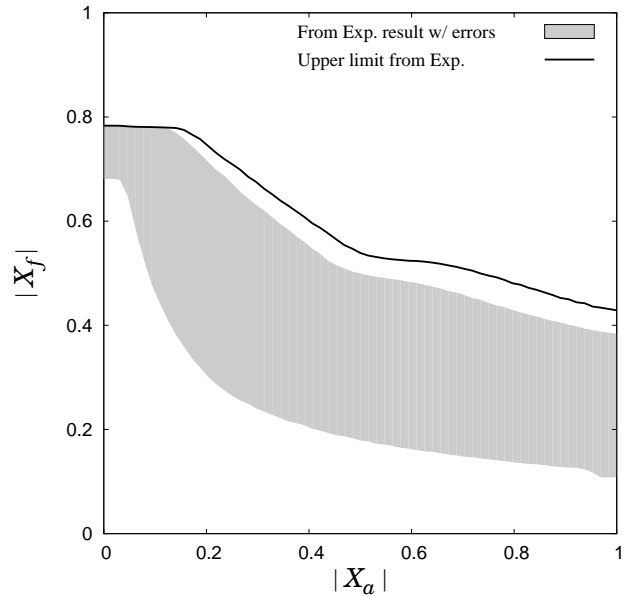}
  \caption{Allowed region for the absolute values of compositeness
    $|X_{a}|$ and $|X_{f}|$ constrained by the experimental value of
    the $a_{0}(980)$-$f_{0}(980)$ mixing intensity~\eqref{eq:xi_exp}
    and \eqref{eq:xi_exp_UL}.  The shaded area corresponds to the
    allowed region within the experimental value~\eqref{eq:xi_exp}
    including errors, and the solid line to the upper limit of the
    experimental value $\xi _{f a}= 1.1 \%$~\eqref{eq:xi_exp_UL}.}
\label{fig:XaXf}
\end{figure}

The experimental value~\eqref{eq:xi_exp} and \eqref{eq:xi_exp_UL}
constrains not only the value of the product $|X_{a} X_{f}|$ but also
the allowed region for $|X_{a}|$ and $|X_{f}|$ in the
$|X_{a}|$-$|X_{f}|$ plane.  Here we show in Fig.~\ref{fig:XaXf} the
allowed region for $|X_{a}|$ and $|X_{f}|$ in the $|X_{a}|$-$|X_{f}|$
plane calculated by sweeping the values of $\bar{g}_{a}$ and
$\bar{g}_{f}$ independently.  In Fig.~\ref{fig:XaXf} the shaded area
corresponds to the allowed region within the experimental
value~\eqref{eq:xi_exp} including errors, and the solid line to the
upper limit from the experimental value~\eqref{eq:xi_exp_UL}, $\xi _{f
  a}= 1.1 \%$.  The region above the solid line in Fig.~\ref{fig:XaXf}
inevitably leads to the mixing intensity $\xi _{f a}>1.1 \%$ and hence
excluded.  As one can see, the experimental value of the mixing
intensity $\xi _{f a}$ does not allow the region of $|X_{a}| \sim
|X_{f}| \sim 1$, thus the statement that both the $a_{0}(980)$ and
$f_{0}(980)$ resonances are simultaneously $K \bar{K}$ molecular
states is questionable.  In fact, this consequence was already implied
in Ref.~\cite{Ablikim:2010aa}, in which the authors showed that the
experimental mixing intensity disfavored the predicted value for
$a_{0}(980)$ and $f_{0}(980)$ as $K \bar{K}$ molecules.  We here note
that conditions that one of the scalar mesons has large degree of the
$K \bar{K}$ molecule, such as $|X_{a}|=0.1$ and $|X_{f}|=0.7$ or
$|X_{a}|=0.8$ and $|X_{f}|=0.3$, are not forbidden by the experimental
value of the mixing intensity.  Especially, the band in
Fig.~\ref{fig:XaXf} show $|X_{f}| \gtrsim 0.3$ regardless of the value
of $|X_{a}|$, which might indicate a nonnegligible degree of the $K
\bar{K}$ molecule for the $f_{0}(980)$ resonance.  Moreover, the
experimental mixing intensity does not disfavor the condition that
both $\azero$ and $\fzero$ have nonnegligible $K \bar{K}$ components
with, for instance, $| X_{a} | = | X_{f} | = 0.4$.

\section{Discussion}
\label{sec:discussion}

So far we have considered the $a_{0}(980)$-$f_{0}(980)$ mixing and
have given a constraint on the $K \bar{K}$ compositeness for the
$a_{0}(980)$ and $f_{0}(980)$ resonances from an established relation
between the $K \bar{K}$ compositeness and the
$a_{0}(980)$-$f_{0}(980)$ mixing intensity.  As a result, the
experimental value of the mixing intensity implies that the
$a_{0}(980)$ and $f_{0}(980)$ resonances cannot be simultaneously $K
\bar{K}$ molecular states.  The $a_{0}(980)$-$f_{0}(980)$ mixing,
however, cannot answer the question that whether one of $a_{0}(980)$
and $f_{0}(980)$ is a $K \bar{K}$ molecular state or not.  In order to
solve this problem experimentally, we have to call for other
experimental data on these resonances.

One possible way to determine the structure of each scalar meson is to
evaluate the compositeness.  Actually Eq.~\eqref{eq:Xi} indicates that
compositeness for each resonance can be evaluated from the pole
position and the coupling constant as the residue at the pole
position.  From this point of view, in Sec.~\ref{sec:KKcomp} we have
calculated the $K \bar{K}$ compositeness of the scalar mesons $\azero$
and $\fzero$ described with the Flatte parameters in
Table~\ref{tab:1}.  From the results shown in Table~\ref{tab:4} and in
Fig.~\ref{fig:KKcomp_Flatte}, we have found that two of the parameter
sets for $f_{0}(980)$ imply large absolute value of the $K \bar{K}$
component with, say $| X_{f} |>0.6$, and the parameters for $\azero$
lead to small but nonnegligible $K \bar{K}$ compositeness $| X_{a} |
\sim 0.2$.  Nevertheless, in order to conclude the structure of the
scalar mesons more strictly, we have to determine the pole positions
and coupling constants more precisely in experiments.  Especially it
will be important to fit the $K \bar{K}$ spectrum as well as the $\pi
\eta$/$\pi \pi$ spectrum, which will lead to a small error of the $K
\bar{K}$ compositeness coming from the $K \bar{K}$ coupling constant,
as done in the BES analysis in Ref.~\cite{Ablikim:2004wn}.

Another approach to determine the structure of the scalar mesons is to
measure their spatial size, since a hadronic molecule can be a
spatially extended object due to the absence of strong quark confining
force.  In fact, the spatial size of exotic hadron candidates was
theoretically measured in a meson-meson and meson-baryon scatterings
in, {\it e.g.}, Refs.~\cite{Sekihara:2010uz, Sekihara:2012xp,
  Sekihara:2008qk, Albaladejo:2012te}, and it was found that
$f_{0}(980)$ and $\Lambda (1405)$, whose $K \bar{K}$ and $\bar{K} N$
compositeness are close to unity, respectively, has spatial size
exceeding largely the typical hadronic scale $\lesssim 0.8 \fm$.

The internal structure of the scalar mesons could be also investigated
by the $\phi$ radiative decays into $f_{0}(980)$ and $a_{0}(980)$,
since $\phi \to f_{0}(980) \gamma$ and $a_{0}(980) \gamma$ are
electric dipole decays and hence the widths should reflect spatial
sizes of the scalar mesons~\cite{Close:1992ay, Oller:1998ia,
  Oller:2002na, Palomar:2003rb, Kalashnikova:2004ta, Achasov:2006cq,
  Kalashnikova:2007dn, Achasov:2008wy}.  The couplings of the scalar
mesons to two photons are also sensitive to the structure of the
scalar mesons and have been investigated in, {\it e.g.},
Refs.~\cite{Oller:1997yg, Mori:2006jj, Oller:2007sh, Uehara:2009cf,
  Achasov:2012ph}. The measurements of the $\phi$ radiative decays and
the two-photon couplings support the multiquark picture for the scalar
mesons $f_{0}(980)$ and $a_{0}(980)$~\cite{Achasov:2012ph}.

In addition, high-energy reactions will be useful for determining
internal structure of exotic hadron candidates since quarks and gluons
are appropriate degrees of freedom at high energies. In this context,
the generalized parton distributions and the generalized distribution
amplitudes (GDAs) can be used to clarify internal configurations of
exotic hadrons, especially $f_{0}(980)$ and
$a_{0}(980)$~\cite{Kawamura:2013wfa} by the GDAs in two-photon
reactions [$\gamma \gamma^{*} \to A \bar{A}$, $A = f_{0}(980), \,
a_{0}(980)$].  Next, asymptotic scaling behavior of the production
cross sections can be a guide to determine internal quark
configurations of the exotic hadrons, such as $\Lambda (1405)$, due to
the constituent-counting rule in perturbative
QCD~\cite{Kawamura:2013iia}.  Fragmentation functions of exotic
hadrons could also provide a clue for finding their internal
configurations by using characteristic differences between favored and
disfavored fragmentations~\cite{Hirai:2007ww}, which could be measured
at KEKB.  Finally, the possibility to extract the hadron structure
from the production yield in heavy ion collisions~\cite{Cho:2010db,
  Cho:2011ew} is also interesting, since one can distinguish hadronic
molecules, compact exotic states, and ordinary quark configurations
from the production yield.

\section{Conclusion}
\label{sec:conclusion}

In this study we investigated the structure of the $a_{0}(980)$ and
$f_{0}(980)$ resonances with the $a_{0}(980)$-$f_{0}(980)$ mixing.
Since the $a_{0}(980)$-$f_{0}(980)$ mixing takes place through the
difference of the thresholds of the charged and neutral $K \bar{K}$
pairs, the mixing should be sensitive to the $K \bar{K}$ components
inside the scalar mesons.  Actually the $a_{0}(980)$-$K \bar{K}$ and
$f_{0}(980)$-$K \bar{K}$ coupling constants reflect the $K \bar{K}$
structure of the $a_{0}(980)$ and $f_{0}(980)$ resonances,
respectively, and the mixing amplitude is proportional to the two
coupling constants.  The key quantity to connect the $a_{0}(980)$- and
$f_{0}(980)$-$K \bar{K}$ coupling constants to their structure is
compositeness, which is defined as the two-body composite part of the
normalization of the total wave function and corresponds to the amount
of two-body states composing resonances as well as bound states.

The $a_{0}(980)$-$f_{0}(980)$ mixing intensity was defined in the same
manner as the analysis by the BES experiment in
Ref.~\cite{Ablikim:2010aa}, where the ratio of two branching fractions
of $J/\psi$, $J/\psi \to \phi f_{0}(980) \to \phi a_{0}(980) \to \phi
\pi \eta$ to $J/\psi \to \phi f_{0}(980) \to \phi \pi \pi$, was
evaluated.  For the $a_{0}(980) \leftrightarrow f_{0}(980)$ mixing
amplitude, we employed three Feynman diagrams; the leading-order
contribution from the $K^{+} K^{-}$ and $K^{0} \bar{K}^{0}$ loops, sum
of which converges and becomes model independent except for the
coupling constants, and a subleading contribution of a soft photon
exchange in the $K^{+} K^{-}$ loop.  We took the Flatte
parametrization for the $a_{0}(980)$ and $f_{0}(980)$ propagators.  In
this construction, when we appropriately fix the parameters of
$a_{0}(980)$ and $f_{0}(980)$ including the $a_{0}(980)$- and
$f_{0}(980)$-$K \bar{K}$ coupling constants, we can calculate the
$a_{0}(980)$-$f_{0}(980)$ mixing intensity $\xi _{f a}$.  By using the
existing Flatte parameter sets from experimental analyses, we found
that two thirds of the combinations of the Flatte parameter sets
reproduce the experimental value with the errors in
Ref.~\cite{Ablikim:2010aa} while only four combinations exceed the
experimental upper limit of the mixing intensity.

From the same Flatte parameters we also calculated the $K \bar{K}$
compositeness for $a_{0}(980)$ and $f_{0}(980)$, $X_{a}$ and $X_{f}$,
respectively.  Although the compositeness with the correct
normalization becomes complex in general for resonance states, we 
found that two of the Flatte parameter sets for $f_{0}(980)$ give
large absolute value of the $K \bar{K}$ component with, say $| X_{f}
|>0.6$, and the parameters for $\azero$ lead to small but
nonnegligible $K \bar{K}$ compositeness $| X_{a} | \sim 0.2$.

Next, combining the two results on the \afmix mixing intensity and on
their $K \bar{K}$ compositeness from the existing Flatte parameters,
we found a relation between the mixing intensity and the the absolute
value of the product of the compositeness, $| X_{a} X_{f} |$, from
experiments.  This relation was confirmed by generating the values of
the $a_{0}(980)$- and $f_{0}(980)$-$K \bar{K}$ coupling constants
randomly, which are responsible for both the mixing intensity and the
$K \bar{K}$ compositeness.  As a result, we found an upper limit of
allowed $|X_{a} X_{f}|$ for each value of $\xi _{f a}$, which behaves
like $|X_{a} X_{f}|_{\rm upper~limit} \sim \xi _{f a}$.  Especially
the result suggests that a small mixing intensity $\xi _{f a}$
directly indicates a small value of $| X_{a} X_{f} |$.  Then, by using
the $a_{0}(980)$-$f_{0}(980)$ mixing intensity recently observed in
the BES experiment, we constrained the allowed region of the $K
\bar{K}$ compositeness $|X_{a}|$ and $|X_{f}|$ in the
$|X_{a}|$-$|X_{f}|$ plane.  We found that the region $|X_{a}| \sim
|X_{f}| \sim 1$ is not preferred, which implies that the $a_{0}(980)$
and $f_{0}(980)$ resonances cannot be simultaneously $K \bar{K}$
molecular states.  However, the analysis does not rule out
possibilities that one of the scalar mesons has large degree of the $K
\bar{K}$ molecule.  Especially, we obtained $|X_{f}| \gtrsim 0.3$
regardless of the value of $|X_{a}|$, which might indicate a
nonnegligible degree of the $K \bar{K}$ molecule for the $f_{0}(980)$
resonance.

\begin{acknowledgments}

  The authors acknowledge T.~Hyodo for fruitful discussions on the
  compositeness.  The authors also thank to E.~Oset for his comment on
  the \afmix mixing phenomenon.
  This work was partly supported by the MEXT KAKENHI Grant Number
  25105010.

\end{acknowledgments}

\end{document}